\documentclass{cmslatex}
\usepackage{latexsym, amssymb, enumerate, amsmath}

    \usepackage{graphics}
\renewcommand{\theequation}{\arabic{section}.\arabic{equation}}

\newtheorem{thm}{Theorem}[section]

\newtheorem{remark}[thm]{Remark}
\usepackage{graphicx}
\DeclareMathOperator{\diff}{d}

\def\MM#1{\boldsymbol{#1}}

\newcommand{\dt}[1]{\diff#1}
\newcommand{\eqnref}[1]{(\ref{#1})}
\newcommand{\eps}{\epsilon}
\newcommand{\cK}{\mathcal{K}}
\newcommand{\cL}{\mathcal{L}}
\newcommand{\R}{\mathbb{R}}
\newcommand{\E}{\mathbb{E}}
\newcommand{\T}{\mathbb{T}}
\newcommand{\epsr}{\sqrt{\epsilon}}

\newcommand{\rato}{\frac{1}{\eps}}
\newcommand{\ratt}{\frac{1}{\eps^2}}
\newcommand{\rath}{\frac{1}{\epsr}}

\newcommand{\remove}[1]{}

\normalsize

\allowdisplaybreaks

\begin{document}

\title{Estimating eddy diffusivities from noisy Lagrangian
observations
}
\author{C.J. Cotter 
\thanks {Department of Aeronautics, Imperial College London, London SW7 2AZ,
UK (colin.cotter@imperial.ac.uk).}
\and G.A. PAVLIOTIS \thanks {Department of Mathematics, Imperial College London, London SW7 2AZ,
UK (g.pavliotis@imperial.ac.uk).}}



\pagestyle{myheadings} \markboth{Estimating Eddy Diffusivities from Noisy Lagrangian Observations}{C.J. COTTER AND G.A. PAVLIOTIS}

\maketitle

\begin{abstract}
  The problem of estimating the eddy diffusivity from Lagrangian
  observations in the presence of measurement error is studied in this
  paper. We consider a class of incompressible velocity fields for
  which is can be rigorously proved that the small scale dynamics can
  be parameterised in terms of an eddy diffusivity tensor. We show, by
  means of analysis and numerical experiments, that subsampling of the
  data is necessary for the accurate estimation of the eddy
  diffusivity. The optimal sampling rate depends on the detailed
  properties of the velocity field. Furthermore, we show that
  averaging over the data only marginally reduces the bias of the
  estimator due to the multiscale structure of the problem, but that
  it does significantly reduce the effect of observation error.
\end{abstract}


\section{Introduction}
\label{sec:intro}
Many phenomena in the physical sciences involve a multitude of
characteristic temporal and spatial scales. In most cases it is not
only impossible to study the behavior of the phenomenon at all scales,
but it is also unnecessary, since usually one is interested in the
evolution of a few variables which describe the dynamics at large
scales. It is, therefore, important to develop systematic methods for
deriving simplified coarse grained models that capture the essential
features of the systems at long scales, while accurately
parameterising the small scales. In recent years it has become clear
that the use of data, together with coarse graining procedures, is
essential for the accurate parameterisation of small scales
\cite{GKS04,CVE06a,CVE06b,IKDS07,HS08}. The aim of this paper is to study
problems of this form in the context of transport of passive tracers.

We are particularly motivated by the challenge of using Lagrangian float data to
inform the design  of  subgrid mixing schemes for
advected tracers in ocean models. The vast amount of Lagrangian float
data available (for example, the ARGO project has 3000 floats in
current operation \cite{ARGO06}) presents the opportunity to develop
data-driven model reduction techniques. Lagrangian data are
particularly suitable for statistical studies of the transport of
passively advected substances in the ocean, with the
simplest statistical description of transport phenomena provided by
the average concentration of a passive tracer.

In this paper, we assume that the Lagrangian trajectories are given by
the following stochastic differential equation:
\begin{equation}\label{e:lagrange}
\dot{x} = v(x,t) + \sqrt{2 \kappa} \dot{W}.
\end{equation}
Here $x(t) \in \R^d$ represents the Lagrangian path, $v(x,t)$ is a
(prescribed) incompressible velocity field, $\kappa$ is the
small-scale diffusivity and $W(t)$ denotes standard Brownian motion in
$\R^d$. More sophisticated models have been proposed for
oceanographic applications, for example
\cite{BeMc02,BeMc03,Pi02,Wi05,Gr+95}.

We wish to extract the coarse grained (large length scale and long
time scale) dynamics of solutions of equation (\ref{e:lagrange}).  For
a wide class of velocity fields (deterministic space-time periodic,
Gaussian random fields \emph{etc.}), it is well known
\cite{kramer,PavlSt08, lions} that, at sufficiently long length and
time scales, the dynamics of~\eqref{e:lagrange} becomes Brownian and
can be described by the {\bf eddy diffusivity tensor}.  More
precisely, it is possible to prove that
solutions of~\eqref{e:lagrange} converge, under the diffusive rescaling and
assuming that the velocity field has zero mean, to an effective Brownian
motion 
\begin{equation}\label{e:weak_limit}
  \lim_{\eps \rightarrow 0} \eps x(t/\eps^2) = \sqrt{2 \cK} W(t),
\end{equation}
weakly on $C([0,T]; \R^d)$, where $W(t)$ is a standard Brownian motion
on $\R^d$ and $\cK$ denotes the eddy (effective) diffusivity
tensor. The tensor $\cK$ represents the effective diffusivity caused
by the interaction of molecular diffusion with the transport
properties of $v$.  

Consequently,
at large length scales and long time scales the dynamics of the
passive tracer is governed by an equation of the form
\begin{equation}\label{e:coarse_grained}
\dot{X} = \sqrt{2 \cK} \dot{W}.
\end{equation}
It is quite often the case (in designing subgrid mixing schemes for
example) that we only wish to calculate the eddy diffusivity, rather
than the detailed properties of the velocity field $v(x,t)$ at all
scales. It is then necessary to estimate the eddy diffusivity of a
passive tracer from Lagrangian observations. In this paper we address
precisely this issue: given a Lagrangian trajectory which is
consistent with~\eqref{e:lagrange} in the presence of observation
noise, how can we estimate the eddy diffusivity $\cK$? This problem has been
studied quite extensively over the last few 
years~\cite{sb:emfde2,haf:eredd,haf:eredd2,sb:emfde,mv:otsms}.

More generally,
we might also want to estimate other coarse grained quantities such as
the effective drift, or we might want to consider a space dependent
eddy diffusivity. This is a challenging problem in statistical
inference: data sampled from~\eqref{e:lagrange} is only consistent
with~\eqref{e:coarse_grained} at sufficiently large scales. In other
words, the difficulty stems from the fact that the
model~\eqref{e:coarse_grained} used for fitting the data is the wrong
model, apart from the large scale part of the data.  Furthermore, we
do not know \emph{a priori} the length and time scales on which the
coarse grained model~\eqref{e:coarse_grained} is valid. On the other
hand, we can perform statistical inference in a fully parametric
setting for~\eqref{e:coarse_grained}, since only the eddy diffusivity
needs to be estimated; statistical inference for~\eqref{e:lagrange}
would require the non-parametric estimation of the velocity field
$v(x,t)$~\cite{CDRS09}. Parameter estimation for diffusion processes
under misspecified or incorrect models has been studied in the
statistics literature~\cite[Sec 2.6.1]{Kut04}.

The problem of parameter estimation for a model that is incompatible
with the available data at small scales was studied in~\cite{PavlSt06,
  PapPavSt08,PavlPokStu08} for a class of fast-slow systems of SDEs
for which the existence of a coarse grained equation for the slow
variables can be proved rigorously. In these papers, parameter
estimation for the {\bf averaging problem}
\begin{subequations}
\label{e:av_intro}
\begin{eqnarray}\label{e:av_intro_x}
\frac{dx}{dt} &=& f_1(x,y)+\alpha_0(x,y)\frac{dU}{dt}+\alpha_1(x,y)\frac{dV}{dt}, \\
\frac{dy}{dt} &=& \rato g_0(x,y)+\rath \beta(x,y)\frac{dV}{dt};
\end{eqnarray}
\end{subequations}
as well as for the {\bf homogenization problem}
\begin{subequations}
\label{e:hom_intro}
\begin{eqnarray}\label{e:hom_intro_x}
\frac{dx}{dt} &=& \rato
f_0(x,y)+f_1(x,y)+\alpha_0(x,y)\frac{dU}{dt}+\alpha_1(x,y)\frac{dV}{dt}, \\
\frac{dy}{dt} & =& \ratt g_0(x,y)+\rato g_1(x,y)+\rato \beta(x,y)\frac{dV}{dt}.
\end{eqnarray}
\end{subequations}
was studied. In both cases the goal was to fit data obtained from~\eqref{e:av_intro_x} or~\eqref{e:hom_intro_x} to the coarse grained equation
\begin{equation}\label{e:coarse_intro}
\frac{d X}{d t} = F(X; \theta) + K(X) \frac{d W}{d t},
\end{equation}
which describes the dynamics of the slow variable $x(t)$ in the limit
as $\eps \rightarrow 0$. In the aforementioned papers, it was assumed
that the vector field $F(X;\theta)$ depends on a set of parameters
$\theta$ that we want to estimate using data taken from either the
averaging or the homogenization problem. For the homogenization
problem it was shown in~\cite{PapPavSt08} that the maximum likelihood
estimator is asymptotically biased. In particular, it is necessary to
subsample at an appropriate rate in order to estimate the parameters
$\theta$ accurately. Similar issues were investigated for the thermal
motion of a particle in a multiscale potential~\cite{PavlSt06}. It was
shown that subsampling is necessary for the accurate estimation of the
drift and diffusion coefficients.

Related issues have been studied in the field of econometrics. In this
context, the question is how to accurately estimate the integrated
stochastic volatility when market microstructure noise (\emph{i.e.}
additive white noise) is present. It was shown in~\cite{AitMykZha05a,
  AitMykZha05b} that subsampling reduces the bias in the estimator. It
was also shown that subsampling combined with averaging and an
appropriate de-biasing step can lead to an accurate and efficient
estimator for the integrated stochastic volatility.

In this paper we will study the problem of estimating the eddy
diffusivity from noisy Lagrangian observations:
\[
y_{t_j} = x_{t_j} + \theta \eps_{t_j}, \quad j=1, \dots N,
\]
where $\{x_0,x_1,\ldots,x_N\}$ is a set of samples from a trajectory
consistent with equation~\eqref{e:lagrange}, $\eps_{t_j}$ are
independent $\mathcal{N}(0,1)$ random variables modelling the
observation error, and $\theta$ measures the strength of the
observation error.

We will consider time-independent spatially-periodic incompressible
velocity fields as well as spatially-periodic velocity fields that are
modulated in time by a time-periodic function, or a Gaussian
process. In all of these cases, the rescaled trajectory converges
weakly to a Brownian motion~\eqref{e:weak_limit} (see
\cite[Ch. 13]{PavlSt08}). The eddy diffusivity depends in a highly
nonlinear way on the properties of the velocity field $v(x,t)$. It can
be shown (for the class of velocity fields considered in this paper)
that the eddy diffusivity $\cK$ satisfies the upper and lower bounds (we use
the notation $\cK^{\xi} = \langle \xi, \cK \xi \rangle$ where $\xi$ is
an arbitrary vector in $\R^d$)~\cite{AvelMajda91}
\begin{equation}\label{e:bounds}
\kappa \leq \cK^{\xi} \leq \frac{C}{\kappa},
\end{equation}
for $\kappa$ sufficiently small and some positive constant $C$.  We
will consider the physically interesting regime $\kappa \ll 1$.

As an eddy diffusivity estimator we will use the quadratic variation
\begin{equation}\label{e:quad_variation}
\cK_{N,\delta} = \frac{1}{2 N \delta} \sum_{n=0}^{N-1} \big( x_{n+1}- x_n
\big) \otimes \big( x_{n+1}- x_n \big),
\end{equation}
where $N$ is the number of observations which we assume to be
equidistant, with the distance between two subsequent observations
being $\delta$, and $T = N \delta$.  It is well known \cite{BasRao80}
that, for an SDE of the form~\eqref{e:lagrange}, we have the
convergence result
\begin{equation}\label{e:kappa_lim}
\lim_{N \rightarrow \infty} \sum_{j=0}^{N-1} \big( x_{(j+1) T 2^{-N}}
- x_{j T 2^{-N}} \big) \otimes \big( x_{(j+1) T 2^{-N}}
- x_{j T 2^{-N}} \big) = 2 \kappa I T, \quad \mbox{a.s.}
\end{equation}
where $I$ denotes the unit matrix. If we write
equation~\eqref{e:kappa_lim} with $x_i$ replaced by $y_i$, the
quadratic variation diverges in the limit as $N\to \infty$ due to the
observation error.  In view of the bounds~\eqref{e:bounds}, it becomes
clear that that the estimator $\cK_{N,\delta}$ underestimates the
value of the eddy diffusivity in this limit. In particular, when the
eddy diffusivity scales like $\kappa^{-1}$ the
estimator~\eqref{e:quad_variation} can underestimate the eddy
diffusivity by several orders of magnitude.

The above suggests that in order to be able to estimate the eddy
diffusivity from Lagrangian data, subsampling at an appropriate rate
is necessary. However, it is not clear \emph{a priori} what the
sampling rate should be. Roughly speaking, we need to look at the data
at the scale for which the coarse grained
description~\eqref{e:coarse_grained} is valid. The estimation of this
time scale is a difficult dynamical question that has been addressed
only partially ~\cite{Fann01, HP07}. The {\it diffusive time}, the
time that it takes for the Lagrangian particle to reach the asymptotic
regime described by a Brownian motion with diffusion matrix $\cK$
depends crucially on the streamline topology and is related to the
scaling of the eddy diffusivity with $\kappa$. Clearly we have two
sources of error: measurement error, and the error in the estimation
of parameters from reduced models using data from the full dynamics
which we refer to as the {\bf multiscale error}. The multiscale error
is precisely due to the fact that the reduced model is incompatible
with the data at small scales.

In this paper we study the small $\kappa$ asymptotics of the quadratic
variation~\eqref{e:quad_variation}.  We show, by means of rigorous
analysis and numerical experiments, that, unless we subsample at an
appropriate rate, we cannot estimate the eddy diffusivity from the
quadratic variation, due to the multiscale error. Additionally, we
show that for smooth time-independent spatially periodic velocity
fields, the scaling of the optimal sampling rate with $\kappa$
depends on the detailed properties of the velocity field. Our analysis
is based on standard limit theorems for stochastic processes, together
with careful study of a Poisson equation posed on the unit torus.

From the point of view of statistics, it is clearly not optimal to
simply ignore most of the available data by subsampling\footnote{We
  remark, however, that the small scale data that we ignore are highly
  correlated and it is not clear how much additional information they
  contain about the eddy diffusivity.}. It is natural, therefore, to
try to use all data through averaging. We experiment with two
different types of averaging: box averaging (computing the quadratic
variation using local averages), and shift averaging (which is related
to the moving averaging method of statistics). We show by means of
numerical experiments, that shift averaging significantly reduces the
effects of observation error, but only marginally reduces the
multiscale error. On the other hand, box averaging increases the bias
of the estimator.


We emphasize that the setting in which we are working is related to
but different from the problems studied in~\cite{PavlSt06,
  PapPavSt08,PavlPokStu08}. In particular, we do not assume \emph{a
  priori} that we have scale separation and that we know the value of
the parameter $\eps$ which measures the degree of scale
separation. Rather, the scale separation is induced by the dynamics
of~\eqref{e:lagrange}. The time scale at which the coarse grained
description is valid is essentially what we need to estimate, since
this provides us with information about the appropriate sampling rate.
For completeness, we also consider the rescaled
problem~\eqref{e:rescaled} below. The rescaled problem and the
original one are equivalent under space-time rescaling for
time-independent velocity fields. However, from the point of view of
estimating the eddy diffusivity they lead to different problems. When
using the quadratic variation to estimate the eddy diffusivity in
equation~\eqref{e:lagrange} we actually study the small $\kappa$
limit, whereas for the rescaled problem we study the limit of infinite
scale separation while keeping $\kappa$ fixed.

The rest of the paper is organized as follows. In
Section~\ref{sec:resc} we we study the problem of estimating the eddy
diffusivity using Lagrangian observations from the rescaled
equation. In Section~\ref{sec:small_kappa} we study the same problem
for the unscaled equation~\eqref{e:lagrange}, and we also study the
effect of observation error on the estimator. In
Section~\ref{sec:numerics} we develop estimators for the eddy
diffusivity which are based on a combination of subsampling with
averaging and present numerical results for various types of
two-dimensional velocity fields. Summary and conclusions are presented
in Section~\ref{sec:conclusions}. Some technical results are included
in the appendices.

%
%
\section{The Rescaled Problem}
\label{sec:resc}

We consider the equation for the rescaled process 
\[
x^\eps(t) = \eps x(t/\eps^2),
\]
given by
\begin{equation}\label{e:rescaled}
\frac{d x}{d t} = \frac{1}{\eps} v \left(\frac{x}{\eps} \right) + \sqrt{2
\kappa} \dot{W},
\end{equation}
where we have dropped the superscript $\eps$ for notational simplicity.. Our goal is to estimate the eddy diffusivity using data
from~\eqref{e:rescaled}, in the parameter regime $\eps \ll 1$ and for $\kappa$
fixed. In particular, we want to find how the sampling
rate should scale with $\eps$ for the accurate estimation of the eddy diffusivity
using the quadratic variation estimator. The main result of this section
is that, provided that the sampling rate is in between the two characteristic
time scales $1$ and $\eps$ of the problem, then the estimator~\eqref{e:quad_variation}
is asymptotically unbiased, in the limit as $\epsilon\to 0.$

We assume that the velocity field
is smooth, divergence-free, mean zero and $1-$periodic, \emph{i.e.}
periodic with period $1$ in each Cartesian direction. Under these
assumptions, the solution to~\eqref{e:rescaled} converges weakly on
$C([0,T]; \R^d)$ to $X$, as $\eps \rightarrow 0$, the solution of
$$
\frac{d X}{d t} =\sqrt{2 \cK} \frac{d W}{d t},
$$
as $\epsilon\to 0$. The eddy diffusivity is given by the formula
\begin{equation}\label{e:eddy_diff}
\cK = \kappa I + \kappa \int_{\T^d} \nabla_z \chi(z) \otimes \nabla_z \chi(z) \, dz
\end{equation}
where the vector field $\chi(z)$ is the solution of the PDE
\begin{equation}\label{e:cell}
-\cL_0 \chi(z) = v(z)
\end{equation}
on $\T^d$ with periodic boundary conditions, and where $\cL_0$ is the
generator of the Markov process $z$ on $\T^d$:
\begin{equation}\label{e:z_process}
\frac{d z}{d t} = v(z) + \sqrt{2 \kappa} \frac{d W}{d t},
\end{equation}
\emph{i.e.}
$$
\cL_0 = v(z) \cdot \nabla_z + \kappa \Delta_z,
$$
with periodic boundary conditions.  We refer to~\cite{PavlSt08} for
the derivation of this result.

Now let $\cK^\xi = \xi \cdot \cK \xi$ where $\xi \in \R^d$
arbitrary. From~\eqref{e:eddy_diff}
it easily follows that
\begin{equation}\label{e:deff_xi}
\cK^\xi =  \kappa \int_{\T^d} |\xi + \nabla_z \chi^\xi|^2 \, dz,
\end{equation}
where $\chi^\xi = \chi \cdot \xi$.
Let $\cK^{\xi}_{N, \delta}$ be the quadratic variation
along the direction $\xi$:
\begin{equation}\label{e:qv_xi}
  \cK^{\xi}_{N,
    \delta} = \frac{1}{2 N \delta} \sum_{n=0}^{N-1} \big( x_{n+1}^{\xi}
  - x^{\xi}_n \big)^2,
\end{equation}
where $x^{\xi}_n = x (n \delta) \cdot \xi$. 

Our goal is to find how the sampling rate $\delta$ should be chosen so
that we can estimate the component of the eddy
diffusivity~\eqnref{e:eddy_diff} along the direction $\xi$
using~\eqref{e:qv_xi}. The following theorem states that the estimator
converges in $L^2$ to the eddy diffusivity in the limit $\epsilon \to
0$, $N\to \infty$, with $T$ fixed.
\begin{theorem}\label{thm:rescaled}
  Let $v(z)$ be a smooth, divergence-free, mean zero, 1-periodic
  vector field and assume that the process $z$ defined
  in~\eqref{e:z_process} is stationary. Then
\begin{equation}
\E |\cK_{N, \delta}^\xi - \cK^\xi |^2  \leq  \frac{C}{N} + C \big(\eps^4 \delta^{-2} + \eps^3 \delta^{-3/2}+  \eps^2 \delta^{-1} + \eps \delta^{-1/2} \big).
\end{equation}
In particular, when $\delta = \eps^{\alpha}, \, \alpha \in (0,2)$, we have
$$
\lim_{N \rightarrow +\infty}\lim_{\eps \rightarrow 0} \E|
\cK^{\xi}_{N, \delta} - \cK^{\xi} |^2 =0,
$$
for $N\delta = T$ fixed (i.e. $N \sim \eps^{-\alpha}$).
\end{theorem}

\begin{remark}
The scaling of the optimal sampling rate with $\eps$, $\delta \sim \eps^{\alpha},
 $ with $\alpha \in (0,2)$ appears to be sharp and it is expected on intuitive
 grounds, since one would expect that the optimal sampling rate should be
 in between the two characteristic time scales of the problem $1$ and $\eps^2$.
\end{remark}

\begin{remark}
The stationarity assumption on $z$ can be removed since even when $z$ starts with arbitrary initial conditions its law converges exponentially fast to the invariant measure of the process which is the Lebesgue measure on $\T^d$. We refer to~\cite{bhatta_2} for the details.
\end{remark}

For the proof of this theorem we will need the following lemma.

\begin{lemma}\label{lem:rescaled}
Let $v(z)$ be a smooth, divergence-free, mean zero, 1-periodic vector field and assume that the process $z$ defined in~\eqref{e:z_process} is stationary. Then
$$
|\E \cK^{\xi}_{N, \delta} - \cK^{\xi} | \leq C \big(\eps^2 \delta^{-1} + \eps \delta^{-1/2} \big).
$$
In particular, when $\delta = \eps^{\alpha}, \, \alpha \in (0,2)$ we
have
$$
\lim_{\eps \rightarrow 0}|\E \cK^{\xi}_{N, \delta} - \cK^{\xi} | =0.
$$
\end{lemma}

\begin{remark}
Notice that in order for the expectation of the quadratic variation to converge to the eddy diffusivity it is not necessary to take the limit $N \rightarrow \infty$. Of course,
in order to keep $N \delta = T$ fixed we need to take $N \sim \delta^{-1} =
\kappa^{-\alpha}$.
\end{remark}

\emph{Proof of Lemma \ref{lem:rescaled}}
  With the help of the auxiliary process $z = x/\eps \in \T^d$,
  equation~\eqref{e:rescaled} can be rewritten as a system of SDEs:
\begin{subequations}\label{e:rescaled_system}
\begin{equation}
\frac{d x}{d t} = \frac{1}{\eps} v \left(z \right) + \sqrt{2
\kappa} \dot{W}.
\end{equation}
\begin{equation}
\frac{d z}{d t} = \frac{1}{\eps^2} v \left(z \right) + \sqrt{\frac{2
\kappa}{\eps^2}} \dot{W}.
\end{equation}
\end{subequations}
The generator of the Markov process $\{x(t), \, z(t) \}$ is
\begin{eqnarray*}
\cL^\eps & =& \frac{1}{\eps^2} \big(v(x)\cdot \nabla_z + \kappa \Delta_z \big)
+ \frac{1}{\eps} \big(v(x)\cdot \nabla_z + 2 \kappa \nabla_x \cdot \nabla_z  \big) + \kappa \Delta_x \\ & =: &  \frac{1}{\eps^2} \cL_0
+ \frac{1}{\eps} \cL_1 + \cL_2.
\end{eqnarray*}
Let $\chi^\xi(z)$ denote the solution of the Poisson equation
$$
-\cL_0 \chi^\xi = v \cdot \xi =:v^\xi(z)
$$
on $\T^d$ with periodic boundary conditions. From standard elliptic
PDE theory we have that $\xi^{\xi} \in C^{\infty}(\T^d)$. Hence, we
can apply It\^{o}'s formula to $\chi^\xi$ and use the fact that
$\chi^\xi$ is independent of $x$ to obtain
\begin{eqnarray*}
d \chi^\xi &=& \left( \frac{1}{\eps^2} \cL_0 \chi^\xi + \frac{1}{\eps} \cL_1 \chi^\xi + \cL_2 \chi^\xi \right)\, dt + \frac{\sqrt{2 \kappa}}{\eps} \nabla_y
\chi^\xi \cdot d W \\ & = & -\frac{1}{\eps^2} v^\xi (z)\, dt + \frac{\sqrt{2 \kappa}}{\eps} \nabla_z \chi^\xi \cdot d W.
\end{eqnarray*}
Consequently:
$$
\frac{1}{\eps} \int_{n \delta}^{(n+1)\delta} v^\xi (z_s) \, ds = - \eps \big(
\chi^\xi(z_{n+1}) - \chi^\xi(z_{n}) \big) + \sqrt{2 \kappa}\int_{n \delta}^{(n+1)\delta} \nabla_z \chi^\xi \cdot d W.
$$
Thus:
\begin{eqnarray*}
x^\xi_{n+1} - x^\xi_n &=& - \eps \big(
\chi^\xi(z_{n+1}) - \chi^\xi(z_{n}) \big) + \sqrt{2 \kappa}\int_{n \delta}^{(n+1)\delta} \big(\nabla_z \chi^\xi + \xi \big) \cdot d W \\ & =: &   \eps R_n + \sqrt{2} M_n.
\end{eqnarray*}
The quadratic variation becomes
\begin{eqnarray*}
\cK_{N, \delta}^\xi &=& \frac{1}{2 N \delta} \sum_{n=0}^{N-1} \left( \eps^2 R_n^2 + 2 \sqrt{2} \eps R_n M_n + 2 M_n^2 \right).
\end{eqnarray*}
Since we have assumed that $z(t)$ is stationary, we have that
\begin{equation}\label{e:Mn}
\E |M_n|^2 = \kappa \|\xi + \nabla_z \chi^\xi \|^2_{L^2(\T^d)} \delta = \cK^\xi \delta
\end{equation}
from which it follows that
$$
\E \left( \frac{1}{ N \delta} \sum_{n=0}^{N-1}   M_n^2 \right) = K^\xi.
$$
Furthermore, the maximum principle for elliptic PDEs implies that
$$
\E |R_n|^2 \leq C.
$$
We use now the above calculations and Cauchy-Schwarz to obtain
\begin{eqnarray*}
\E \cK_{N, \delta}^\xi - \cK^\xi & = &   \frac{1}{2 N \delta} \sum_{n=0}^{N-1}   \eps^2 \E R_n^2 + 2 \sqrt{2} \eps \E \big( R_n M_n \big) \\ & \leq & C \big( \eps^2 \delta^{-1} +  \eps \delta^{-1/2} \big).
\end{eqnarray*}
\noindent $\square$

\noindent {\it Proof of Theorem~\ref{thm:rescaled}}

From Lemma~\ref{lem:rescaled} we have that
\begin{eqnarray*}
  \E |\cK_{N, \delta}^\xi - \cK^\xi |^2 & = & \E | \cK_{N, \delta}^\xi |^2 - |\cK^\xi|^2 + 2 \cK^\xi \big(\cK^{\xi} - \cK^{\xi}_{N,\delta}  \big) \\ & \leq & \E | \cK_{N, \delta}^\xi |^2 - |\cK^\xi|^2 + C \big( \eps^2 \delta^{-1} +  \eps \delta^{-1/2} \big).
\end{eqnarray*}
Hence, it is sufficient to estimate the difference $E | \cK_{N, \delta}^\xi |^2 - |K^\xi|^2$. Using the notation introduced in the proof of Lemma~\ref{lem:rescaled} we can write
\begin{eqnarray}
|K^{\xi}_{N, \delta}|^2 & = & \frac{1}{4 N^2 \delta^2} \sum_{n=0}^{N-1}\sum_{\ell=0}^{N-1} \left( \eps^2 R_n^2 + 2 \sqrt{2} \eps R_n M_n + 2 M_n^2 \right)  \left( \eps^2 R_{\ell}^2 + 2 \sqrt{2} \eps R_{\ell} M_{\ell} + 2 M_{\ell}^2 \right) \nonumber   \\ & = & \frac{1}{ N^2 \delta^2}\sum_{n=0}^{N-1}\sum_{\ell=0}^{N-1} M_n^2 M_{\ell}^2 + R, \label{e:kxi}
\end{eqnarray}
where
\begin{eqnarray*}
R & = & \frac{1}{4 N^2 \delta^2} \sum_{n=0}^{N-1}\sum_{\ell=0}^{N-1} \Big( \eps^4 R_n^2 R_{\ell}^2 + 4 \sqrt{2} \eps^3 R_n^2 R_{\ell} M_{\ell}
\\ &&+ 4 \eps^2 R_n M_{\ell}^2 + 8 \eps^2 R_n R_{\ell} M_n M_{\ell} +8 \sqrt{2} \eps M_n M_{\ell}^2 R_n \Big) \\ & =:& I + II + III + IV + V.
\end{eqnarray*}
The uniform bound on $\chi^{\xi}$ and its derivatives, bounds on moments of stochastic integrals~\cite{KSh91} and the Cauchy-Schwarz inequality yield the bounds
\begin{eqnarray*}
\E \, I \leq C \eps^4 \delta^{-2}, \quad \E \, II \leq C \eps^3 \delta^{-3/2}, \quad \E \, III \leq C \eps^2 \delta^{-1}, \quad \E \, IV \leq C \eps^2 \delta^{-1}, \quad \E \, V \leq C \eps \delta^{-1/2}.
\end{eqnarray*}
From the above bounds we deduce that
\begin{equation}\label{e:r_estim}
\E \, R \leq C \big(\eps^4 \delta^{-2} + \eps^3 \delta^{-3/2}+  \eps^2 \delta^{-1} + \eps \delta^{-1/2} \big).
\end{equation}
Now we use bounds on moments of stochastic integrals, together with the fact that $\E (M_n M_\ell) = 0$ for $n \neq \ell$ to calculate
\begin{eqnarray*}
\E \left(\frac{1}{ N^2 \delta^2}\sum_{n=0}^{N-1}\sum_{\ell=0}^{N-1} M_n^2 M_{\ell}^2 \right) & = & \frac{1}{ N^2 \delta^2}\sum_{n=0}^{N-1} \E ( M_n^4 ) + \frac{1}{ N^2 \delta^2}\sum_{n=0}^{N-1}\sum_{\ell \neq n} \E ( M_n^2 ) \E( M_{\ell}^2)  \\ & \leq & \frac{C}{N} +   \frac{1}{ N^2 \delta^2}\sum_{n=0}^{N-1}\sum_{\ell \neq n} \E ( M_n^2 ) \E( M_{\ell}^2).
\end{eqnarray*}
On the other hand, from Equation~\eqref{e:Mn} we deduce that
\begin{eqnarray}
\E \left|K^\xi_{N,\delta} \right|^2 & = & \frac{1}{ N^2 \delta^2}\sum_{n=0}^{N-1}\sum_{\ell =0}^{N-1} \E ( M_n^2 ) \E( M_{\ell}^2) \nonumber
\\ & = &
\frac{1}{ N^2 \delta^2}\sum_{n=0}^{N-1}\sum_{\ell \neq n} \E ( M_n^2 ) \E( M_{\ell}^2) + \mathcal{O} \left(\frac{1}{N} \right). \label{e:kxi_delta}
\end{eqnarray}
We combine the above estimates to obtain
\begin{eqnarray*}
\E |\cK_{N, \delta}^\xi - \cK^\xi |^2 & \leq &  \E | \cK_{N, \delta}^\xi |^2 - |K^\xi|^2 + C \big( \eps^2 \delta^{-1} +  \eps \delta^{-1/2} \big) \\ & \leq & \frac{C}{N} + C \big(\eps^4 \delta^{-2} + \eps^3 \delta^{-3/2}+  \eps^2 \delta^{-1} + \eps \delta^{-1/2} \big).
\end{eqnarray*}

\noindent $\square$

%
%

\section{Small $\kappa$ Asymptotics for the Quadratic Variation}
\label{sec:small_kappa}

In this section we consider the original problem
\begin{equation}\label{e:sde}
\frac{d x}{d t} = v(x) + \sqrt{2 \kappa} \frac{d W}{d t}.
\end{equation}
Our goal is to estimate the eddy diffusivity using data
from~\eqref{e:sde}, in the parameter regime $\kappa \ll 1$. In particular, we want to find how the sampling
rate should scale with $\kappa$ for the accurate estimation of the eddy diffusivity
using the quadratic variation estimator. The main result of this section
is that in order for the estimator~\eqref{e:quad_variation} to be asymptotically
unbiased in the limit as $\kappa \rightarrow 0$, it is necessary that the
sampling rate (as well as the number of observations, and hence the time interval of observation) must  scale with $\kappa$ in an appropriate way,
which depends on the detailed properties of the velocity field. In particular,
the optimal sampling rate might become unbounded in the limit as $\kappa
\rightarrow 0$ for flows for which the eddy diffusivity also becomes unbounded
in this limit. Furthermore, our results are not sharp and detailed analysis
is required for each particular flow. In contrast with the rescaled problem
that was studied in the previous section, there doesn't seem to be a simple
intuitive argument to explain the scaling of the optimal sampling rate with $\kappa$, since the longest characteristic time scale of the problem (the
diffusive time scale) needs to be estimated, as a function of $\kappa$.

As in the previous section we are interested in analyzing the quadratic variation along
an arbitrary direction $\xi$ and to calculate the optimal sampling rate in order to be able to estimate the eddy diffusivity from observations. Let $\cK^{\xi}_{N,
\delta}$ be the quadratic variation along the direction $\xi$ is given by Equation~\eqref{e:qv_xi}
\begin{equation}\label{e:qv_unresc}
\cK^{\xi}_{N,
\delta} = \frac{1}{2 N \delta} \sum_{n=0}^{N-1} \big( x_{n+1}^{\xi} - x^{\xi}_n \big)^2
\end{equation}
where $x^{\xi}_n = x (n \delta) \cdot \xi$. The eddy diffusivity along the direction $\xi$ is given by Equation~\eqref{e:deff_xi}
\begin{equation*}
\cK^\xi =  \kappa \int_{\T^d} |\xi + \nabla_z \chi^\xi|^2 \, dz
\end{equation*}
where $\chi^\xi = \chi \cdot \xi$ is the unique mean zero solution of the elliptic PDE
\begin{equation}\label{e:cell_kappa}
- ( v(z) \cdot \nabla_z + \kappa \Delta_z ) \chi^{\xi} = v^\xi
\end{equation}
with periodic boundary conditions on the unit torus.
In order to study the small $\kappa$ asymptotics of the quadratic variation
$\cK_{N, \delta}^\xi$ we need information on the small $\kappa$ asymptotics of $\chi^\xi$,
the solution of~\eqref{e:cell_kappa}. From the PDE~\eqref{e:cell_kappa} and Poincar\'{e}'s
inequality we deduce the bounds
$$
\|\chi^\xi \|_{L^2} \leq C \|\nabla_z \chi^\xi \|_{L^2} \leq \frac{C}{\kappa}.
$$
The precise asymptotic behavior of $\chi^\xi$ in the small $\kappa$
regime depends on the detailed properties of the velocity field
$v(z)$. This difficult problem has been studied quite
extensively~\cite{ConstKiselRyzhZl06, bhatta_1, Fann01, MajMcL93}. In
this paper we will assume that the solution of the cell problem
satisfies the following small-$\kappa$ scaling
\begin{equation}\label{e:kappa_scaling}
\|\chi^\xi \|_{L^p} \sim   \|\nabla_z \chi^\xi \|_{L^p} \sim \kappa^{\alpha},
\quad \alpha \in [-1,0], \quad \kappa \ll 1,
\end{equation}
for $p=2, \, 4$. The notation $ f \sim \kappa^\alpha$ means that there exists
constants $C_+, \, C_-$ so that
$$
C_- \kappa^\alpha \leq f \leq C_+ \kappa^\alpha, \quad \mbox{for} \;\; \kappa \ll 1.
$$
Some examples of flows for which the scaling of $\chi^\xi$, the solution
of~\eqref{e:cell_kappa}, with $\kappa$ is known are:
\begin{enumerate}
\item The two-dimensional shear flow ${\bf v}({\bf x}) = (0, \sin(x))$
~\cite{kramer,MajMcL93}. For this flow we can solve 
the Poisson equation explicitly:
$$
\chi_1(x,y) = 0, \quad \chi_2 (x,y) = -\kappa^{-1} \sin(x)
$$
and, consequently, for all $\kappa > 0$,
$$
\|\chi_2 \|_{L^p} \sim \|\nabla \chi_2 \|_{L^p} \sim \kappa^{-1}.
$$
\item The Taylor-Green flow
$$
{\bf v}(x,y) = \nabla^{\bot} \psi_{TG}(x,y), \, \phi_{TG}(x,y) = \sin(x) \sin(y).
$$
In this case it is not possible to solve~\eqref{e:cell_kappa}. However it is possible
to obtain sharp estimates on the solution of the Poisson equation:
$$
\|\nabla \chi^\xi \|_{L^2} \sim \kappa^{-1/2}, \quad \kappa \ll 1
$$
for all vectors $\xi \in \R^2$. See~\cite{Heinz03} for details. On the other hand, by the
maximum principle we have that $\|\chi \|_{L^2} \leq C$, uniformly in
$\kappa$~\cite{Fann01}.
\item The Childress-Soward flow
$$
{\bf v}(x,y) = \nabla^{\bot} \psi_{CS}(x,y), \, \phi_{CS}(x,y) = \sin(x) \sin(y) +
\lambda\cos(x) \cos(y),
$$
where $\lambda\in[0,1]$. This flow interpolates between the
Taylor-Green flow (for $\delta=0$) and a shear flow (for $\delta=1$).

In this case we have that
$$
\|\chi^{\xi_1} \|_{L^2} \sim \kappa^{-1}, \quad \|\chi^{\xi_2} \|_{L^2} \sim 1,
$$
where $\xi_1 = 1/\sqrt{2} (1,1), \, \xi_2 = 1/\sqrt{2} (-1,1)$. 
See~\cite{childress1,Fann01} for details.
\end{enumerate}
More examples of flows for which the small-$\kappa$ asymptotics of $\chi^\xi$  can be
calculated will be presented in Section~\ref{sec:numerics}.

\begin{remark}
Notice that the above scaling leads to
\begin{equation}\label{e:deff_scaling}
\cK^{\xi} \sim \kappa^{2 \alpha +1},
\end{equation}
which is consistent with~\eqref{e:bounds}, since $2 \alpha +1 \in [-1,1]$.
\end{remark}
\begin{remark}
Of course, the exponent $\alpha$ in~\eqref{e:kappa_scaling} in general depends
in the direction $\xi$ as well as the $L^p$-space, $\alpha =\alpha(\xi,p)$. For simplicity
we will assume that $\alpha$ is independent of $p$. The analysis presented below can be
easily extended to cover the case where $\alpha =\alpha(p)$.
\end{remark}

\subsection{Convergence results}

In this section we prove the following.

\begin{theorem}\label{thm:l2_unrescaled}
Let $v(z)$ be a smooth, divergence-free smooth vector field on $\T^d$.
Assume that the scaling~\ref{e:kappa_scaling} with $p=4$ holds.
Then the following estimate holds
\begin{eqnarray}
\E |\cK^\xi_{N,\delta} - \cK^\xi|^2
 \leq  C \Big(\frac{1}{N} \kappa^{4 \alpha + 2}
+\kappa^{4 \alpha + 1}  \delta^{-1} + \kappa^{4 \alpha} \delta^{-2}
+ \kappa^{4 \alpha + \frac{3}{2}} \delta^{-\frac{1}{2}}
+ \kappa^{4 \alpha +\frac{1}{2}} \delta^{-\frac{3}{2}}  
\Big). \label{e:l2_bound}
\end{eqnarray}
In particular, if $N \sim \kappa^{\zeta}$ with
$\zeta > 4 \alpha +2$ and $\delta \sim \kappa^{\gamma}$ with
$\gamma < \min(4\alpha +1, 2 \alpha,  8 \alpha+1, \frac{8 \alpha}{3} + \frac{1}{3})$. Then
\begin{equation}\label{e:l2_convergence}
\lim_{\kappa \rightarrow 0} \E |\cK_{N,\delta}^\xi - \cK^\xi|^2 =0.
\end{equation}
\end{theorem}

\begin{remark}
Estimate~\eqref{e:l2_bound} is not sharp. See the examples of the steady and
modulated in time shear flows in the next section.
\end{remark}

\begin{remark}
  Notice that $T = N\delta \to \infty$ as $\kappa$ goes to $\infty$,
  and notice that the sampling rate may also have to go to $\infty$
  depending on the value of $\alpha$. This is in constrast to the
  rescaled problem, for which convergence occurs as $\epsilon \to 0$
  with $T$ fixed.
\end{remark}

We first prove the following weak convergence result.

\begin{lemma}\label{lem:expect_unresc}
Let $v(z)$ be a smooth, divergence-free smooth vector field on $\T^d$.
Assume that the scaling~\ref{e:kappa_scaling} with $p=2$ holds
$$
\big| \E \cK^{\xi}_{N,\delta} - \cK^{\xi} \big| \leq C \big(
\kappa^{2 \alpha + \frac{1}{2}} \delta^{-\frac{1}{2}} + \kappa^{2 \alpha} \delta^{-1} \big).
$$
In particular, if $\delta = \kappa^{\gamma}$ with $\gamma < \min(2 \alpha, 4 \alpha+1)$ then
$$
\lim_{\kappa \rightarrow 0} \big| \E \cK^{\xi}_{N,\delta} - \cK^{\xi} \big| = 0.
$$
\end{lemma}

\begin{proof}

We apply It\^{o}'s formula to $\chi^\xi$ to write the increment of the process $x^\xi$ as
\begin{eqnarray}
x^\xi_{n+1} - x^{\xi}_n & = & \sqrt{2 \kappa} \int_{n \delta}^{(n+1) \delta} 
\big( \nabla_z \chi^\xi + \xi \big) \cdot d W - (\chi^\xi(z_{n+1} ) - \chi^\xi(z_{n} ))
\nonumber \\ & =: & \sqrt{2} M_n + R_n, \label{e:decompose}
\end{eqnarray}
where
$$
\langle M_n \rangle = \kappa \int_{n \delta}^{(n+1) \delta} |\nabla_z \chi^\xi + \xi|^2 
\, dz  \quad \mbox{and} \quad   \E \langle M_n \rangle = \delta \cK^\xi.
$$

Upon combining~\eqref{e:qv_xi} and~\eqref{e:decompose} and taking the expectation we obtain
\begin{eqnarray*}
\E \cK^{\xi}_{N,\delta} = \cK^{\xi} + \frac{\sqrt{2}}{N \delta} \sum_{n=0}^{N-1} \E \left( M_n \, R_n \right) + \frac{1}{2 N \delta} \sum_{n=0}^{N-1} \E  R_n^2.
\end{eqnarray*}
We use now~\eqref{e:kappa_scaling} and~\eqref{e:deff_scaling} to deduce that
\begin{eqnarray}
\E \cK^{\xi}_{N,\delta} - \cK^{\xi}   & \leq & \frac{\sqrt{2}}{N \delta} \sum_{n=0}^{N-1} 
(\E M_n^2)^{1/2} \, ( \E R_n^2)^{1/2} + \frac{1}{2 N \delta} \sum_{n=0}^{N-1} \E R_n^2 \nonumber   \\
 & \leq & C \kappa^{2 \alpha + \frac{1}{2}} \delta^{-\frac{1}{2}} + C \kappa^{2 \alpha} 
 \delta^{-1}. \label{e:estimate_l1}
\end{eqnarray}
\end{proof}

\noindent {\it Proof of Theorem~\ref{thm:l2_unrescaled}.}

From Lemma~\ref{lem:expect_unresc} we have that
\begin{equation}\label{e:R_defn}
\E \cK^\xi_{N,\delta} = \cK^\xi + R
\end{equation}
with
\begin{equation}\label{e:R_estimate}
|R| \leq C \big( \kappa^{2 \alpha + \frac{1}{2}} \delta^{-\frac{1}{2}} + 
\kappa^{2 \alpha} \delta^{-1} \big).
\end{equation}
We can write
\begin{equation}\label{e:kxi2}
\E |\cK^\xi_{N,\delta} - \cK^\xi|^2 = \E \big| \cK^\xi_{N,\delta} \big|^2 - 
(\cK^\xi)^2 - 2 R \cK^\xi.
\end{equation}
We introduce the notation
\begin{equation*}
\big| \cK^\xi_{N, \delta} \big|^2 = I^2 + II^2 + III^2 + 2 I \, II + 2 I \, 
III + 2 II \, III
\end{equation*}
with
\begin{eqnarray*}
I = \frac{1}{N \delta} \sum_{n=0}^{N-1} M_n^2, \quad II = \frac{\sqrt{2}}{N \delta} 
\sum_{n=0}^{N-1} M_n R_n, \quad III = \frac{1}{ 2 N \delta} \sum_{n=0}^{N-1} R_n^2.
\end{eqnarray*}
We use~\eqref{e:kxi_delta} to deduce that
$$
\E I^2 = \frac{N-1}{N} |K^\xi|^2 + \frac{1}{N^2 \delta^2} \sum_{n=0}^{N-1} \E M_n^4.
$$
Furthermore,
\begin{eqnarray*}
\E M_n^4  & = & \E \left( \sqrt{\kappa} \int_{n \delta}^{(n+1) \delta} (\xi + 
\nabla_z \chi^\xi) \, dW \right)^4 \\ & \leq & C \kappa^2 \delta^2 \|\xi + 
\nabla_z \chi^\xi \|^4_{L^2(\T^d)}
\end{eqnarray*}
Scaling~\ref{e:kappa_scaling} together with bounds on moments of stochastic integrals
implies that
$$
\E M_n^4 \leq C \kappa^{4 \alpha+2} \delta^2.
$$
We conclude that
$$
\E I^2 \leq |\cK^\xi|^2 + C \frac{1}{N} \kappa^{4 \alpha +2}.
$$
Consequently
$$
\E I^2 \leq C \left(1 + \frac{1}{N} \right) \kappa^{4 \alpha +2}.
$$

From Assumption~\eqref{e:kappa_scaling} we get
$$
\left( \E |R_n|^p \right)^{1/p} \leq C \kappa^\alpha.
$$
Now we have
\begin{eqnarray*}
\E II^2 & = & \E \left(\frac{2}{N^2 \delta^2} \sum_{n=0}^{N-1} \sum_{k=0}^{N-1} 
R_n M_n R_k M_k \right) \\ & \leq & \frac{2}{N^2 \delta^2} \sum_{n=0}^{N-1} 
\sum_{k=0}^{N-1} (\E|R_n|^4)^{1/4} (\E |M_n|^4)^{1/4} (\E|R_k|^4)^{1/4} 
(\E |M_k|^4)^{1/4} \\ & \leq & C \kappa^{4\alpha +1} \delta^{-1}.
\end{eqnarray*}
Similarly,
\begin{eqnarray*}
\E III^2 & = & \E \left(\frac{1}{4 N^2 \delta^2} \sum_{n=0}^{N-1} \sum_{k=0}^{N-1} 
R_n^2 R_k^2 \right) \\ & \leq & C \kappa^{4 \alpha} \delta^{-2}.
\end{eqnarray*}
We use now the Cauchy-Schwarz inequality to obtain the estimates (we use the
fact that $N \geq 1$)

\begin{eqnarray*}
\E (I \, II) & \leq & C \kappa^{4 \alpha +\frac{3}{2}} \delta^{-\frac{1}{2}}, \\
\E (I \, III) & \leq & C \kappa^{4 \alpha + 1} \delta^{-1} , \\
\E (II \, III) & \leq & C \kappa^{4 \alpha + \frac{1}{2}} \delta^{-\frac{3}{2}}.
\end{eqnarray*}
We use all of the above estimates, together with~\eqref{e:kxi2} and 
estimate~\eqref{e:R_estimate}, to obtain estimate~\eqref{e:l2_bound}.

\noindent $\square$


\begin{figure}
\begin{center}
\includegraphics*[width=12cm]{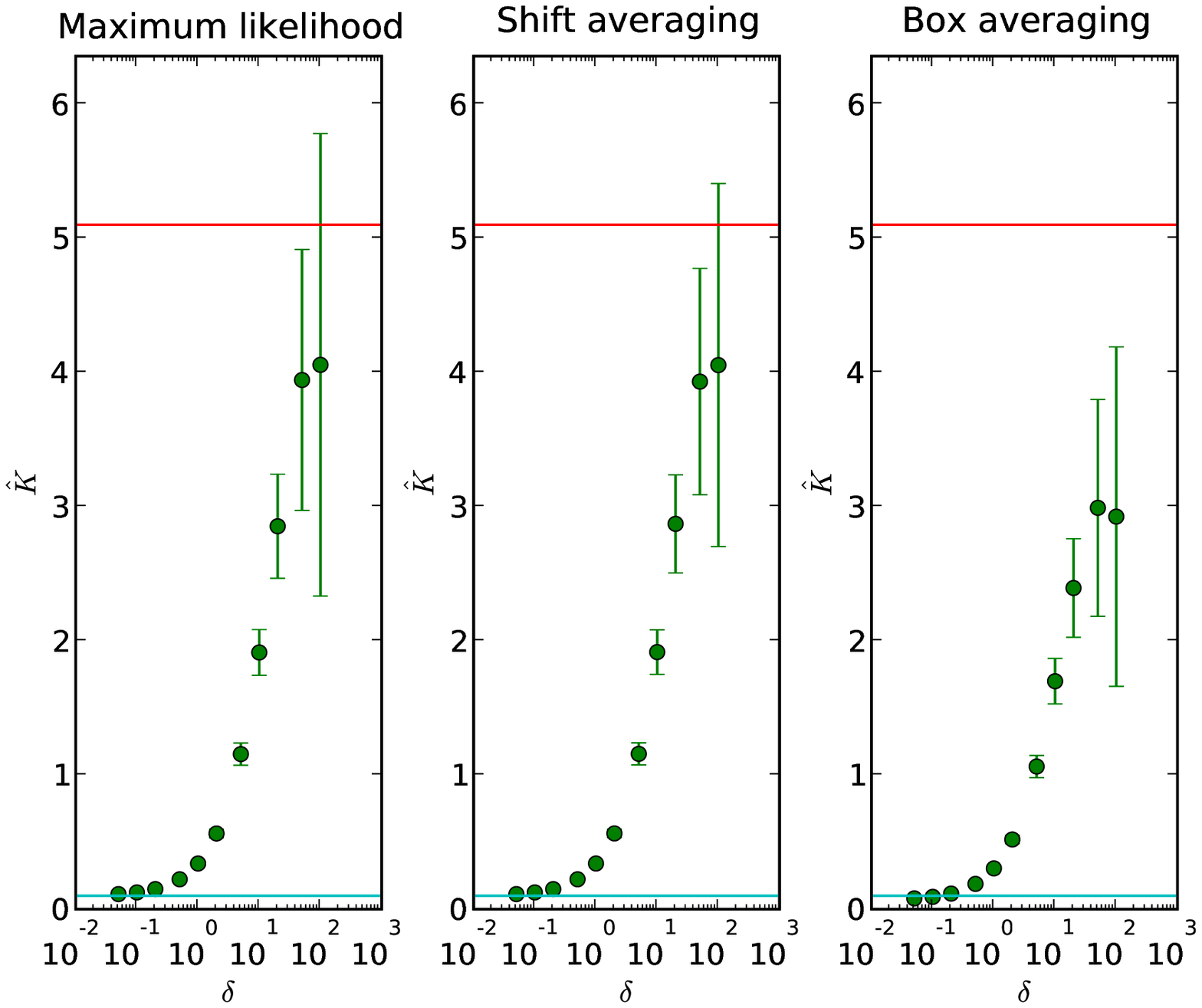}
\end{center}
\caption{\label{shearbars}Figure showing statistics for estimators of
  the eddy diffusivity for the shear flow. The plots show results for
  various values of the subsampling interval $\delta$ from (left) the
  maximum likelihood estimator \eqnref{e:quad_variation}, (centre) the
  shift-averaged estimator \eqnref{shift averaging}, and (right) the
  box-averaged estimator \eqnref{box averaging}.  The plots indicate
  the mean value of the estimators (circular dots), as well as the
  standard deviation (bars) with statistics computed from 1000
  realisations of the Lagrangian trajectory. The correct value
  $\cK=5.1$, and the value of the small-scale diffusivity
  $\kappa=0.1$ are both indicated as horizontal lines.}
\end{figure}

\begin{figure}
\begin{center}
\includegraphics*[width=12cm]{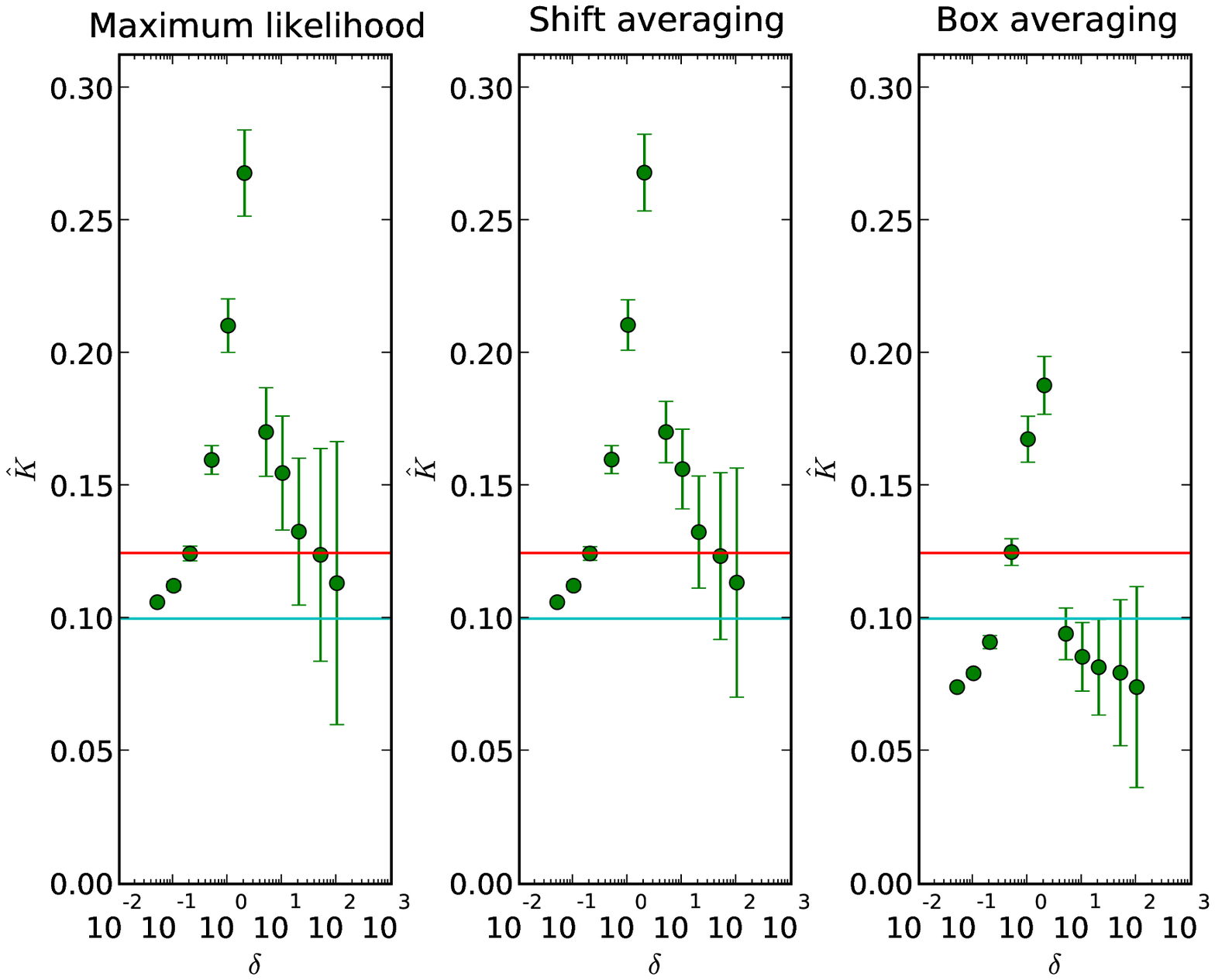}
\end{center}
\caption{\label{modbars}Figure showing statistics for estimators of
  the eddy diffusivity for the periodically-modulated shear flow with
  modulation frequency $\omega=1$. The plots show results for various
  values of the subsampling interval $\delta$ from (left) the maximum
  likelihood estimator \eqnref{e:quad_variation}, (centre) the
  shift-averaged estimator \eqnref{shift averaging}, and (right) the
  box-averaged estimator \eqnref{box averaging}.  The plots indicate
  the mean value of the estimators (circular dots), as well as the
  standard deviation (bars) with statistics computed from 1000
  realisations of the Lagrangian trajectory. The correct value
  $\cK=0.125$ (3 d.p.), and the value of the small-scale
  diffusivity $\kappa=0.1$ are both indicated as horizontal lines.}
\end{figure}

\begin{figure}
\begin{center}
\includegraphics*[width=12cm]{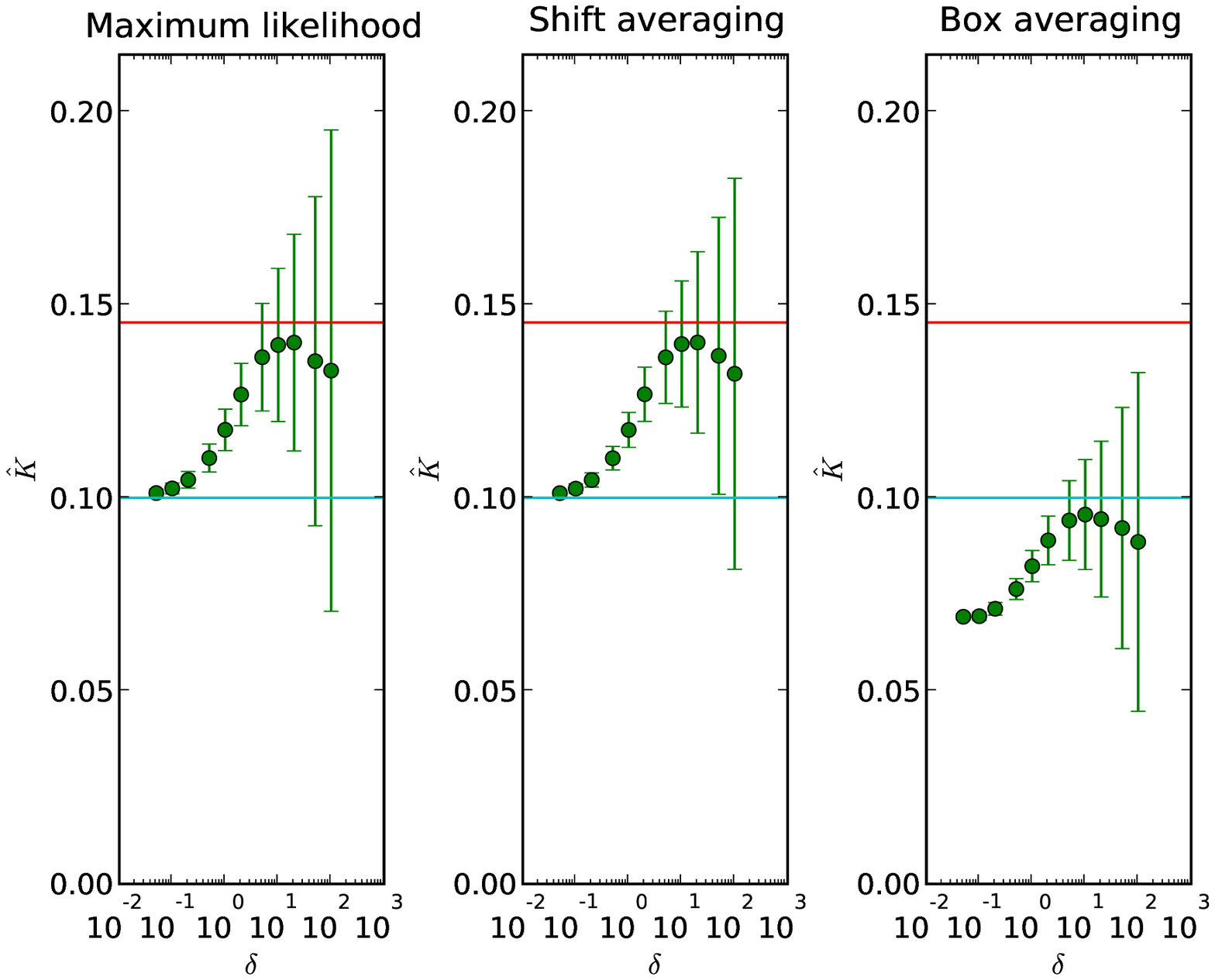}
\end{center}
\caption{\label{OUbars}Figure showing statistics for estimators of
  the eddy diffusivity for the OU-modulated shear flow
  with parameters $\alpha=1$, $\sigma=0.1$. The plots show results for
  various values of the subsampling interval $\delta$ from (left) the
  maximum likelihood estimator \eqnref{e:quad_variation}, (centre) the
  shift-averaged estimator \eqnref{shift averaging}, and (right) the
  box-averaged estimator \eqnref{box averaging}.  The plots indicate
  the mean value of the estimators (circular dots), as well as the
  standard deviation (bars) with statistics computed from 1000
  realisations of the Lagrangian trajectory. The correct value
  $\cK=0.145$ (3 d.p.), and the value of the small-scale diffusivity
  $\kappa=0.1$ are both indicated as horizontal lines.}
\end{figure}

\begin{figure}
\begin{center}
\includegraphics*[width=12cm]{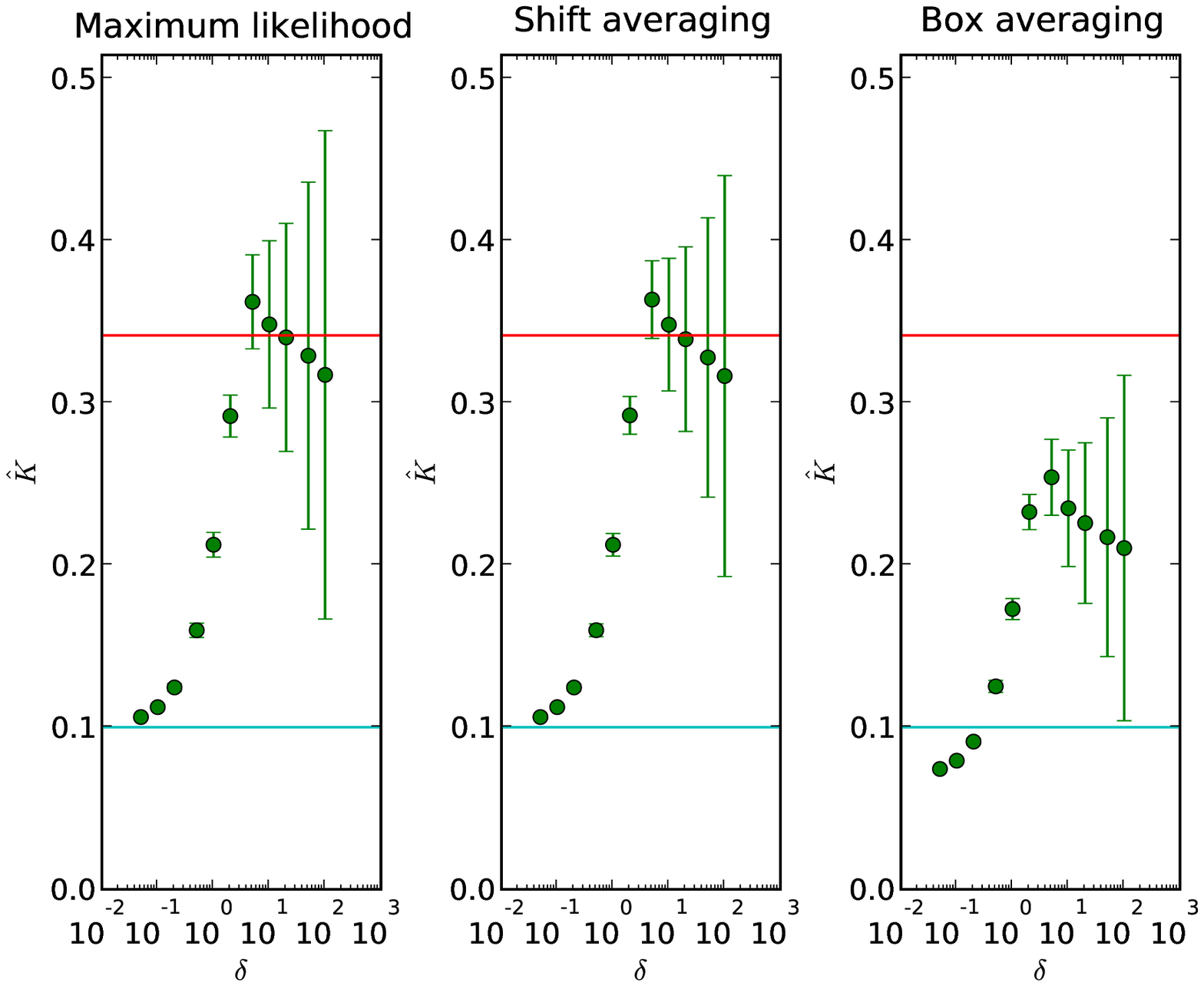}
\end{center}
\caption{\label{TGbars}Figure showing statistics for estimators of the
  eddy diffusivity for the Taylor-Green flow. The plots show results
  for various values of the subsampling interval $\delta$ from (left)
  the maximum likelihood estimator \eqnref{e:quad_variation}, (centre)
  the shift-averaged estimator \eqnref{shift averaging}, and (right)
  the box-averaged estimator \eqnref{box averaging}.  The plots
  indicate the mean value of the estimators (circular dots), as well
  as the standard deviation (bars) with statistics computed from 1000
  realisations of the Lagrangian trajectory. The correct value
  $\cK=0.342$ (3 d.p.), and the value of the small-scale
  diffusivity $\kappa=0.1$ are both indicated as horizontal lines.}
\end{figure}

\begin{figure}
\begin{center}
\includegraphics*[width=12cm]{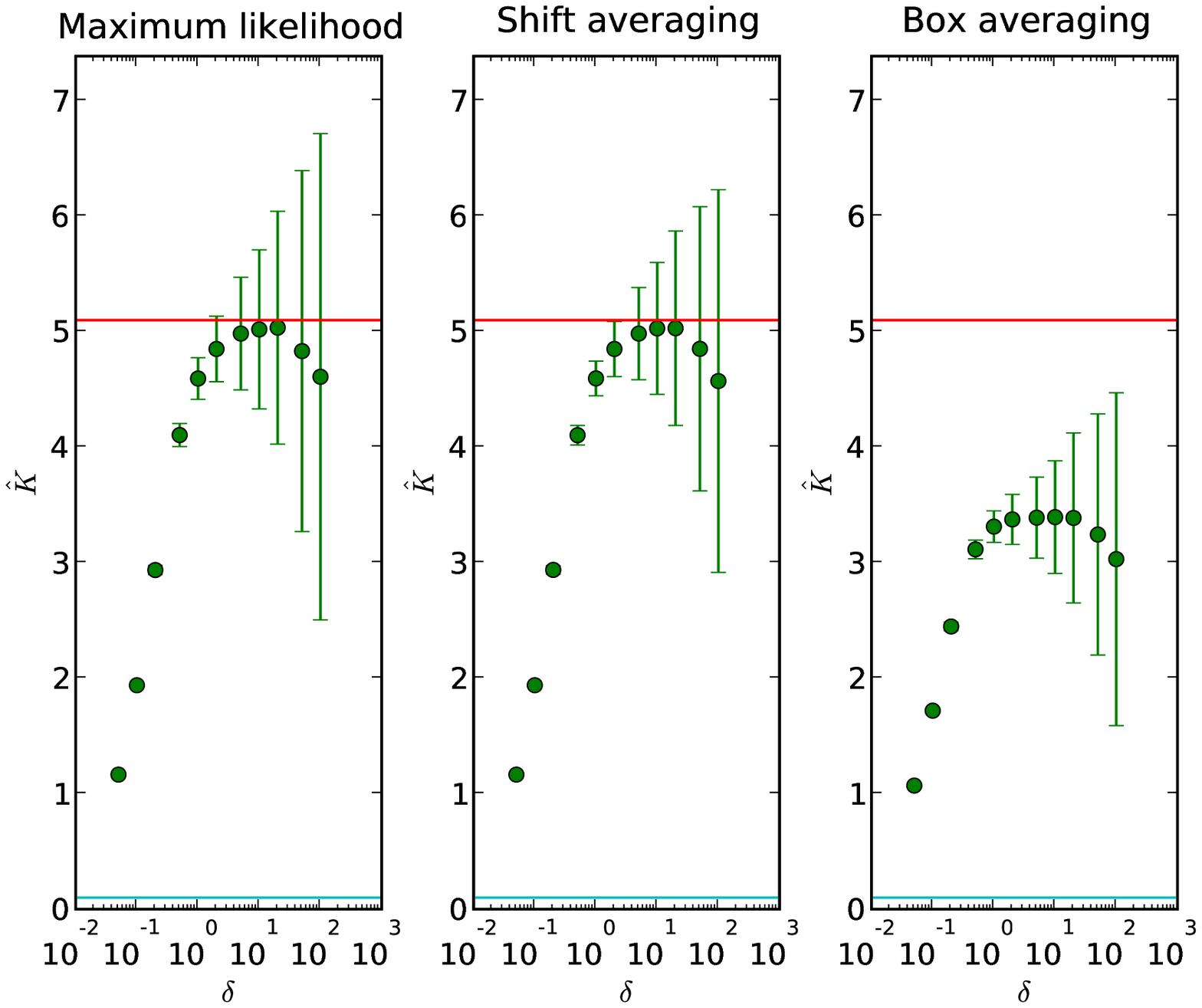}
\end{center}
\caption{
  \label{eps01shearbars} Figure showing statistics for estimators of
  the eddy diffusivity for the shear flow, applied to the rescaled
  problem with $\epsilon=0.1$. The plots show results for various
  values of the subsampling interval $\delta$ from (left) the maximum
  likelihood estimator \eqnref{e:quad_variation}, (center ) the
  shift-averaged estimator \eqnref{shift averaging}, and (right) the
  box-averaged estimator \eqnref{box averaging}.  The plots indicate
  the mean value of the estimators (circular dots), as well as the
  standard deviation (bars) with statistics computed from 1000
  realisations of the Lagrangian trajectory. The correct value
  $\cK=5.1$, and the value of the small-scale diffusivity
  $\kappa=0.1$ are both indicated as horizontal lines.  }
\end{figure}

\begin{figure}
\begin{center}
\includegraphics*[width=12cm]{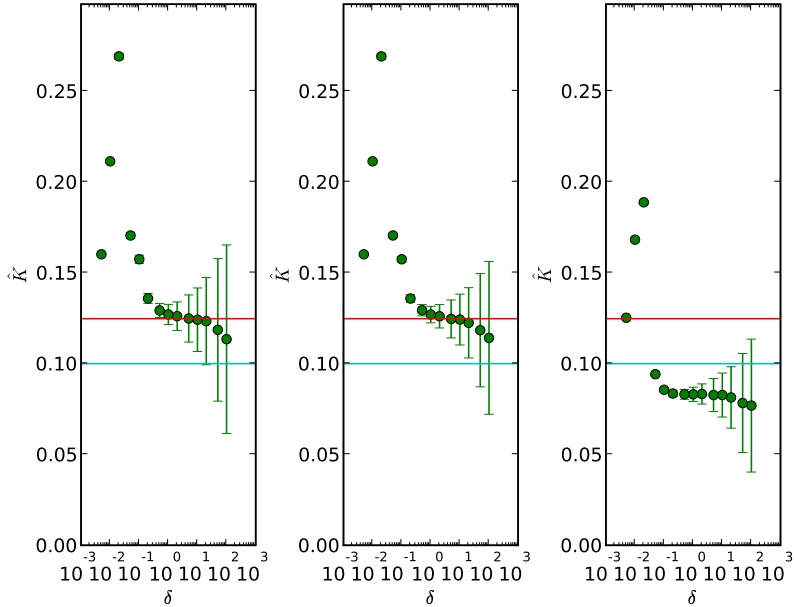}
\end{center}
\caption{\label{eps01modbars}Figure showing statistics for estimators
  of the eddy diffusivity for the periodically-modulated shear flow
  with modulation frequency $\omega=1$, applied to the rescaled
  problem with $\epsilon=0.1$. The plots show results for various
  values of the subsampling interval $\delta$ from (left) the maximum
  likelihood estimator \eqnref{e:quad_variation}, (centre) the
  shift-averaged estimator \eqnref{shift averaging}, and (right) the
  box-averaged estimator \eqnref{box averaging}.  The plots indicate
  the mean value of the estimators (circular dots), as well as the
  standard deviation (bars) with statistics computed from 1000
  realisations of the Lagrangian trajectory. The correct value
  $\cK=0.125$ (3 d.p.), and the value of the small-scale
  diffusivity $\kappa=0.1$ are both indicated as horizontal lines.}
\end{figure}

\begin{figure}
\begin{center}
\includegraphics*[width=12cm]{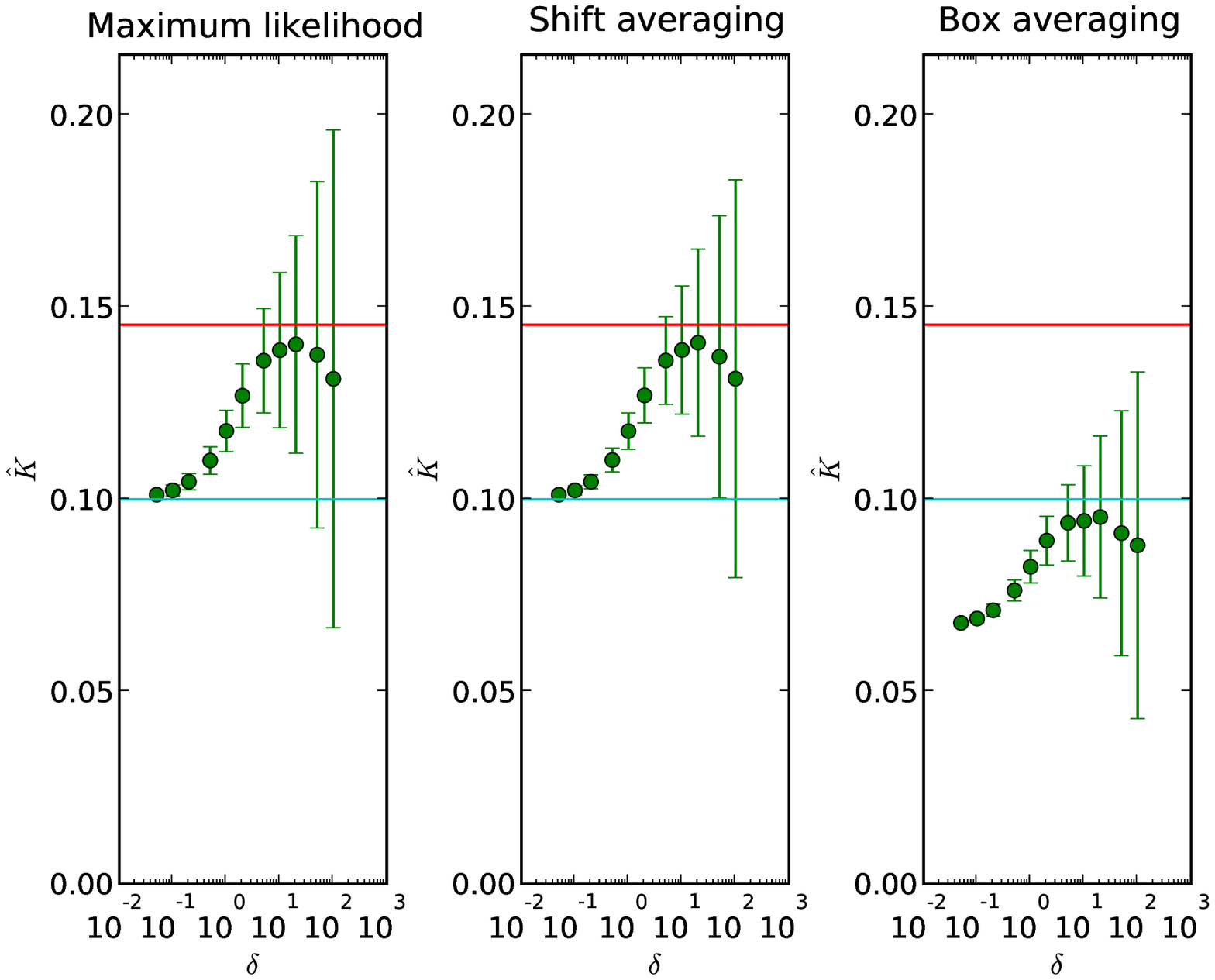}
\end{center}
\caption{\label{eps01OUbars}Figure showing statistics for estimators
  of the eddy diffusivity for the OU-modulated shear flow with
  parameters $\alpha=1$, $\sigma=0.1$, applied to the rescaled problem
  with $\epsilon=0.1$. The plots show results for various values of
  the subsampling interval $\delta$ from (left) the maximum likelihood
  estimator \eqnref{e:quad_variation}, (centre) the shift-averaged
  estimator \eqnref{shift averaging}, and (right) the box-averaged
  estimator \eqnref{box averaging}.  The plots indicate the mean value
  of the estimators (circular dots), as well as the standard deviation
  (bars) with statistics computed from 1000 realisations of the
  Lagrangian trajectory. The correct value $\cK=0.145$ (3 d.p.), and
  the value of the small-scale diffusivity $\kappa=0.1$ are both
  indicated as horizontal lines.}
\end{figure}

\begin{figure}
\begin{center}
\includegraphics*[width=12cm]{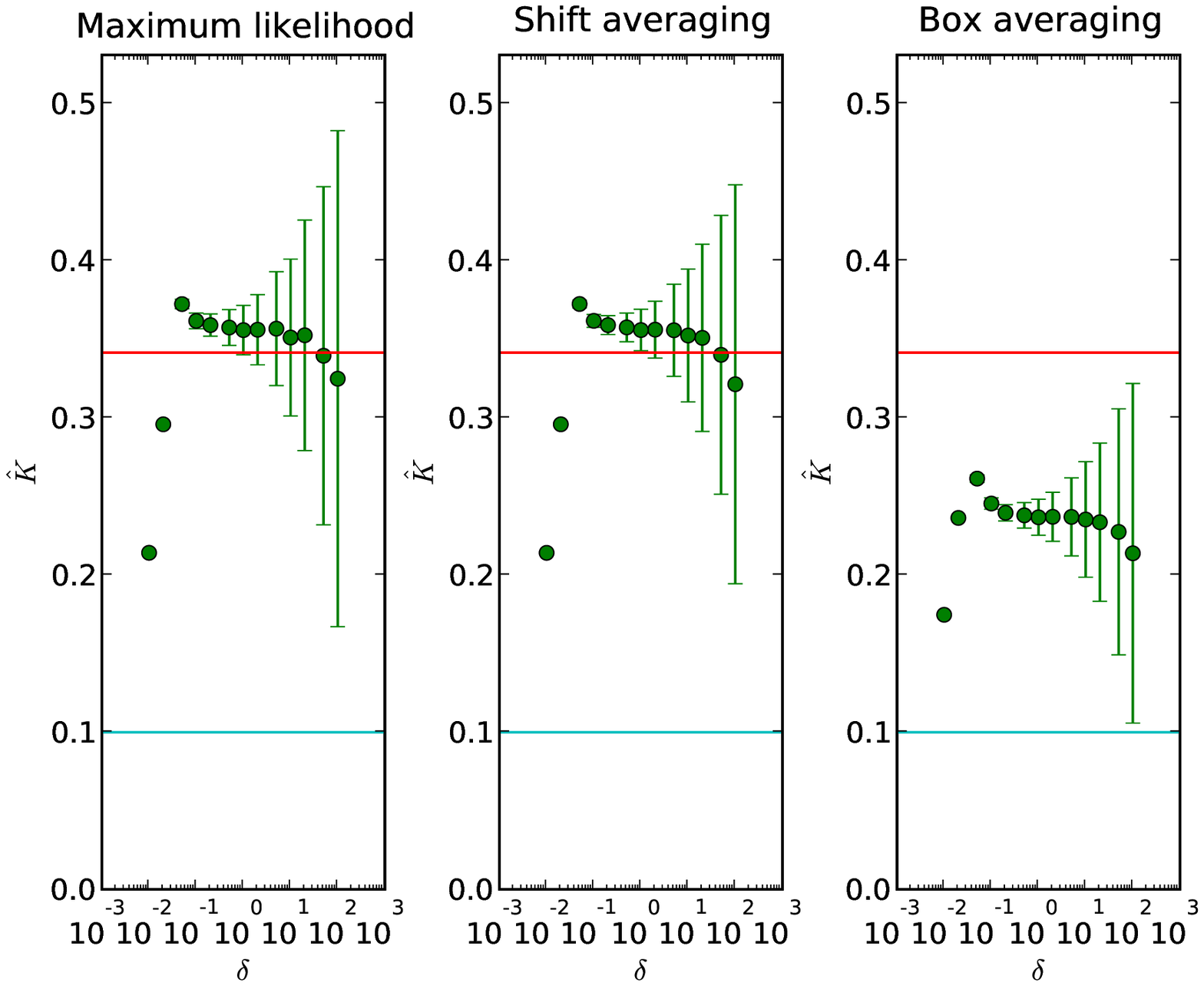}
\end{center}
\caption{\label{eps01TGbars}Figure showing statistics for estimators
  of the eddy diffusivity for the Taylor-Green
  flow, applied to the rescaled problem with $\epsilon=0.1$. The plots
  show results for various values of the subsampling interval $\delta$
  from (left) the maximum likelihood estimator
  \eqnref{e:quad_variation}, (centre) the shift-averaged estimator
  \eqnref{shift averaging}, and (right) the box-averaged estimator
  \eqnref{box averaging}.  The plots indicate the mean value of the
  estimators (circular dots), as well as the standard deviation (bars)
  with statistics computed from 1000 realisations of the Lagrangian
  trajectory. The correct value $\cK=0.342$ (3 d.p.), and the
  value of the small-scale diffusivity $\kappa=0.1$ are both indicated
  as horizontal lines.}
\end{figure}

\begin{figure}
\begin{center}
\includegraphics*[width=12cm]{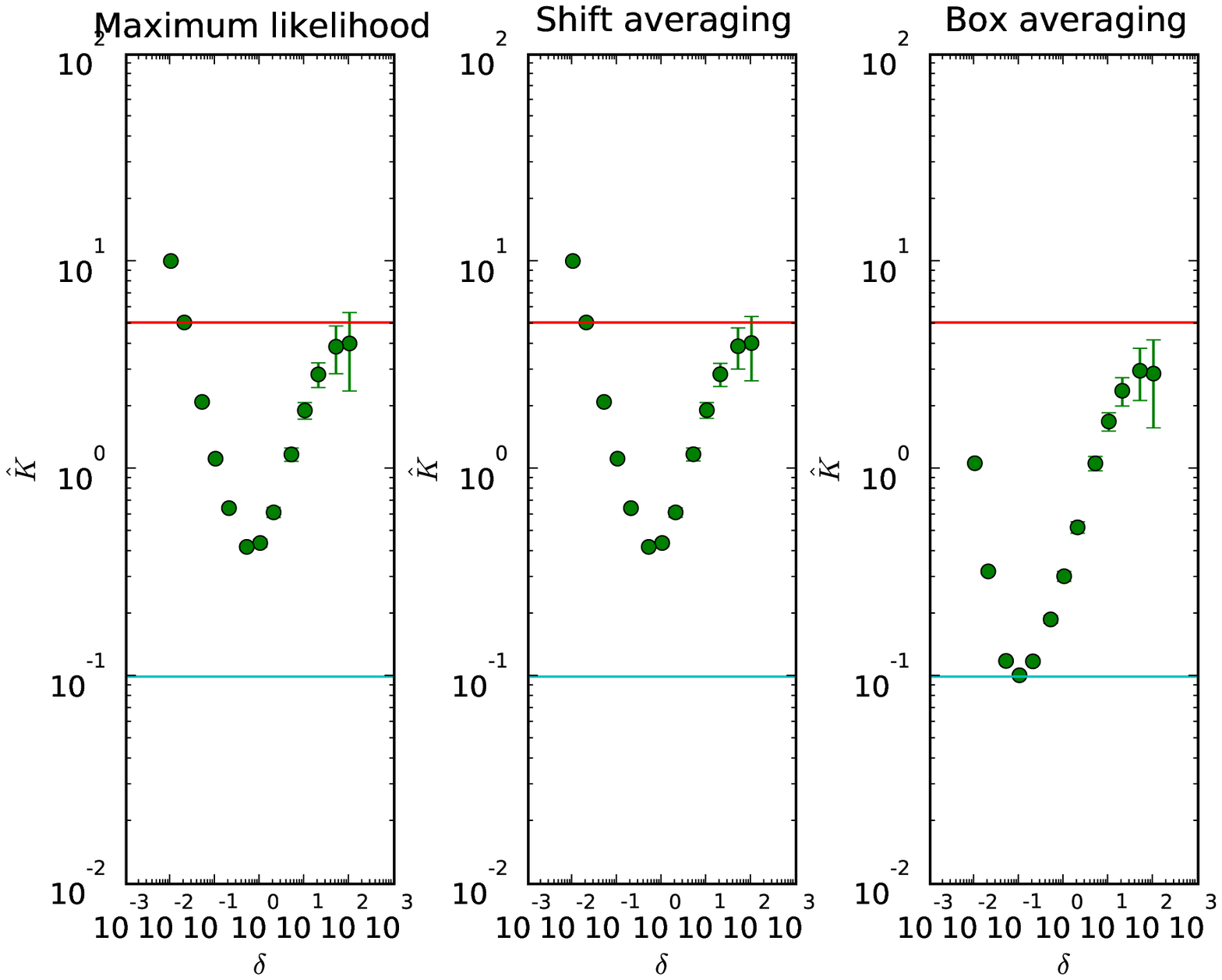}
\end{center}
\caption{
  \label{shearnoise} Figure showing statistics for estimators of
  the eddy diffusivity for the shear flow, where $\mathcal{N}(0,0.1)$
  observation noise has been added. The plots show results for various
  values of the subsampling interval $\delta$ from (left) the maximum
  likelihood estimator \eqnref{e:quad_variation}, (center ) the
  shift-averaged estimator \eqnref{shift averaging}, and (right) the
  box-averaged estimator \eqnref{box averaging}.  The plots indicate
  the mean value of the estimators (circular dots), as well as the
  standard deviation (bars) with statistics computed from 1000
  realisations of the Lagrangian trajectory. The correct value
  $\cK=5.1$, and the value of the small-scale diffusivity
  $\kappa=0.1$ are both indicated as horizontal lines.  }
\end{figure}

  \begin{figure}
  \begin{center}
  \includegraphics*[width=12cm]{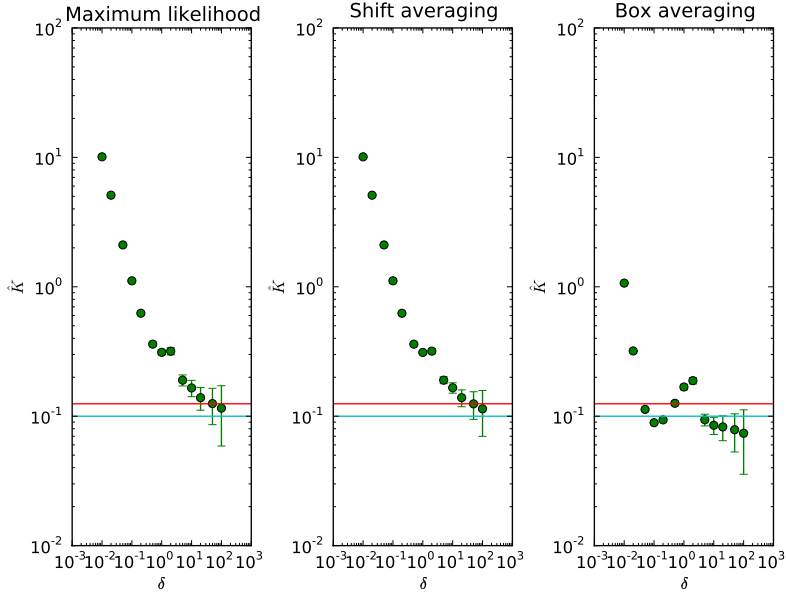}
  \end{center}
  \caption{\label{modnoise} Figure showing statistics for estimators
    of the eddy diffusivity for the periodically-modulated shear flow
    with modulation frequency $\omega=1$, where $\mathcal{N}(0,0.1)$
   observation noise has been added.
  The plots show results for various
    values of the subsampling interval $\delta$ from (left) the maximum
    likelihood estimator \eqnref{e:quad_variation}, (centre) the
    shift-averaged estimator \eqnref{shift averaging}, and (right) the
    box-averaged estimator \eqnref{box averaging}.  The plots indicate
    the mean value of the estimators (circular dots), as well as the
    standard deviation (bars) with statistics computed from 1000
    realisations of the Lagrangian trajectory. The correct value
    $\cK=0.125$ (3 d.p.), and the value of the small-scale
    diffusivity $\kappa=0.1$ are both indicated as horizontal lines.}
  \end{figure}

  \begin{figure}
  \begin{center}
  \includegraphics*[width=12cm]{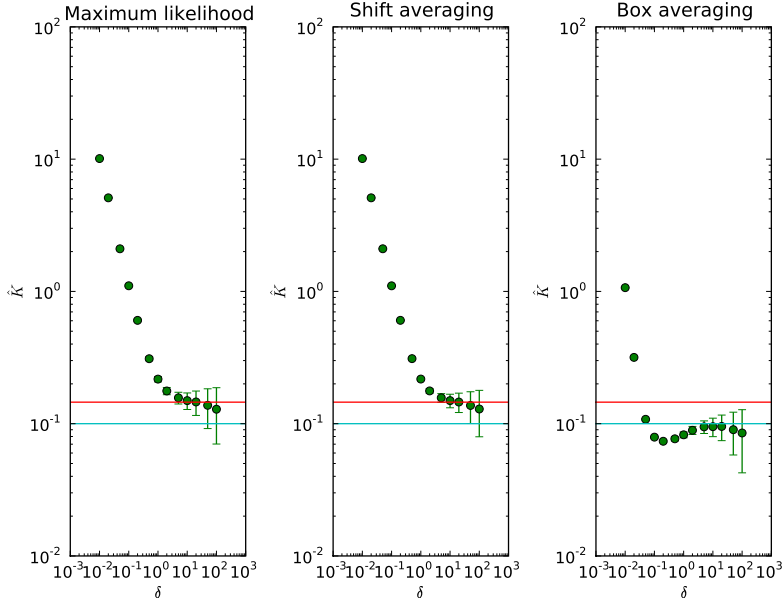}
  \end{center}
  \caption{\label{OUnoise} Figure showing statistics for estimators
    of the eddy diffusivity for the OU-modulated shear flow with
    parameters $\alpha=1$, $\sigma=0.1$,  where $\mathcal{N}(0,0.1)$
   observation noise has been added.
    with $\epsilon=0.1$. The plots show results for various values of
    the subsampling interval $\delta$ from (left) the maximum likelihood
    estimator \eqnref{e:quad_variation}, (centre) the shift-averaged
    estimator \eqnref{shift averaging}, and (right) the box-averaged
    estimator \eqnref{box averaging}.  The plots indicate the mean value
    of the estimators (circular dots), as well as the standard deviation
    (bars) with statistics computed from 1000 realisations of the
    Lagrangian trajectory. The correct value $\cK=0.145$ (3 d.p.), and
    the value of the small-scale diffusivity $\kappa=0.1$ are both
    indicated as horizontal lines.}
  \end{figure}

\begin{figure}
\begin{center}
\includegraphics*[width=12cm]{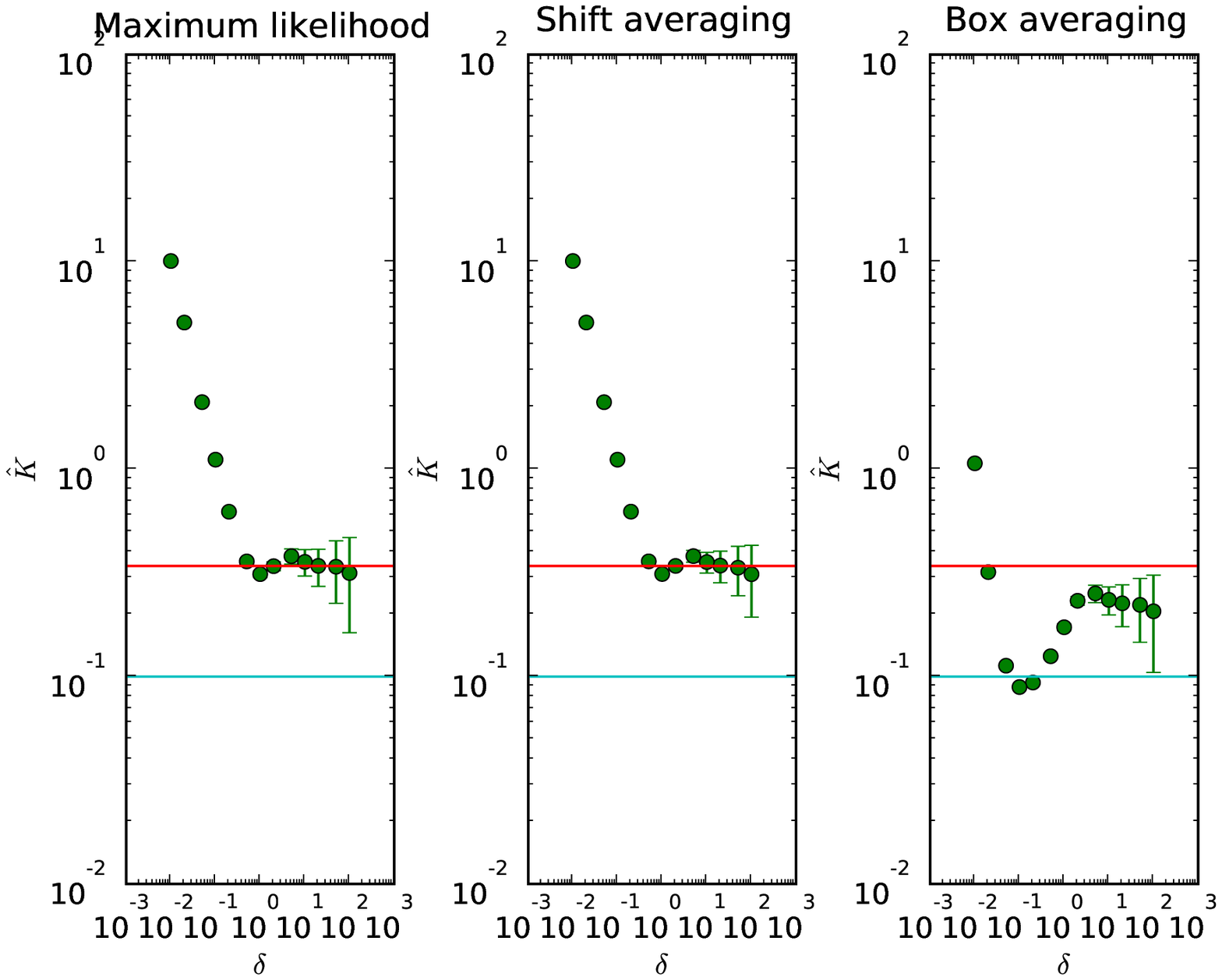}
\end{center}
\caption{\label{TGnoise} Figure showing statistics for estimators of
  the eddy diffusivity for the Taylor-Green flow, where
  $\mathcal{N}(0,0.1)$ observation noise has been added. The plots
  show results for various values of the subsampling interval $\delta$
  from (left) the maximum likelihood estimator
  \eqnref{e:quad_variation}, (centre) the shift-averaged estimator
  \eqnref{shift averaging}, and (right) the box-averaged estimator
  \eqnref{box averaging}.  The plots indicate the mean value of the
  estimators (circular dots), as well as the standard deviation (bars)
  with statistics computed from 1000 realisations of the Lagrangian
  trajectory. The correct value $\cK=0.342$ (3 d.p.), and the value of
  the small-scale diffusivity $\kappa=0.1$ are both indicated as
  horizontal lines.}
\end{figure}


\subsection{The Effect of Observation Error}

In this subsection we study the small $\kappa$ asymptotics of the quadratic
variation in the presence of observation error. More specifically, we assume
that the observed process (along the direction $\xi$) is
\begin{equation}\label{e:noise}
Y^{\xi}_{t_j} = X^{\xi}_{t_j} + \theta \eps^{\xi}_{t_j}, \quad j=1, \dots N.
\end{equation}
The parameter $\theta >0$ measures the strength of the measurement
noise which we model through a collection of i.i.d $\mathcal{N}(0,1)$
random variables $\eps^{\xi}_{t_j}$, which are independent from the
Brownian motion driving the Lagrangian dynamics.  Since the two
sources of noise that appear in the problem are assumed to be
independent, the analysis presented in this section also applies to
Equation~\eqref{e:noise}. In particular, we have that
\begin{equation*}
\E \cK^{\xi}_{N,\delta} (Y_t) = \E \left( \cK^{\xi}_{N,\delta} (X_t)\right) + \frac{\theta^2}{\delta}.
\end{equation*}
In view of estimate~\eqref{e:estimate_l1}, we have that
$$
\left| \E \left( \cK^{\xi}_{N,\delta} (Y_t) \right)  - \cK^{\xi} \right| \leq  C \kappa^{2 \alpha + \frac{1}{2}} \delta^{-\frac{1}{2}} + C \kappa^{2 \alpha} 
 \delta^{-1} + \theta^2 \delta^{-1}.
$$
In particular, if $\delta = \kappa^{\gamma}$ with $\gamma < \min(2 \alpha, 4 \alpha+1, 0)$ then
$$
\lim_{\kappa \rightarrow 0} \big| \E \cK^{\xi}_{N,\delta} - \cK^{\xi} \big| = 0.
$$
We remark that the exponent $\gamma$ is different to the one that appears
in the statement of Lemma~\ref{lem:expect_unresc}, in that it must
 be negative, irrespective of the scaling of the eddy diffusivity with
$\kappa$.

Similarly, in the presence of measurement error, estimate~\eqref{e:l2_bound}
has to be modified. It becomes
\begin{equation}\label{e:unresc_noise}
\E \left|\cK^\xi_{N,\delta} (Y_t) - \cK^\xi \right|^2 = 
\E \left|\cK^\xi_{N,\delta} (X_t) - \cK^\xi \right|^2 + 3 \frac{\theta^4}{\delta^2} + 2 \theta^2 \left(\frac{1}{\delta} + \frac{2}{N
\delta} \right) (\cK^\xi + R),
\end{equation}
where $R$ is defined in equation~\eqref{e:R_defn} and estimated 
in~\eqref{e:R_estimate}. We can then use Theorem~\ref{thm:l2_unrescaled} to bound the first term on the right hand side of equation~\eqref{e:unresc_noise}.
Clearly, we require that $\delta \rightarrow \infty$ for the additional terms
(which are due to the measurement error) to vanish.
%
%
\subsection{The Two-Dimensional Shear Flow}
In this section we present some results for a particular class of flows for
which we can compute the quadratic variation explicitly. The purpose of this
is to show that the results obtained in Theorem~\ref{thm:l2_unrescaled} are
not sharp.

For two-dimensional flows of the form
\begin{equation}\label{e:shear_gen}
v(x,y,t) = (0, \eta(t) \sin(x)),
\end{equation}
where $\eta (t)$ can be either a constant, a periodic function or a
stochastic process, we can calculate explicitly the statistics of the
quadratic variation of the Lagrangian trajectories~\cite{ma:mmert, McL98, kramer}. In the appendix it is shown that for $\eta(t) \equiv 1$,
the quadratic variation along the direction of the shear is
\begin{eqnarray}\label{e:quad_shear}
\mathbb{E} \cK_{N, \delta} 
& = & \cK + \frac{1}{2\kappa^2\delta}(e^{-\kappa\delta}
-1) + \frac{1}{4\kappa^2T}
\left(
\frac{2}{3}e^{-\kappa\delta} - \frac{1}{6}e^{-4\kappa\delta}-\frac{1}{2}
\right)\frac{1-e^{-4\kappa T}}{1-e^{-4\kappa\delta}},
\end{eqnarray}
where $T=N\delta$ and the effective diffusivity is
$$
\cK = \kappa + \frac{1}{2 \kappa}.
$$
From the above formula we immediately deduce that
\begin{eqnarray*}
\lim_{\kappa \rightarrow 0}\mathbb{E} \left[ \cK_{N, \delta}-\cK \right] =   0 
\end{eqnarray*}
provided that 
\begin{equation}\label{e:scaling_shear}
\delta=\kappa^{-2-\epsilon}.
\end{equation} 
for $\eps >0$, arbitrary. Furthermore, when~\eqref{e:scaling_shear} holds,
we have that
\begin{equation}\label{e:conv_rate}
\lim_{\kappa \rightarrow 0} \kappa^{-\eps} \left( \mathbb{E}  \cK_{N, \delta}-\cK \right) = -\frac{1}{2} - \frac{1}{8 N},
\end{equation}
the convergence being exponential in $\kappa$.

It is also possible to calculate $\E |\cK_{N, \delta} - \cK|^2$. In particular,
we have that
\begin{eqnarray}
\nonumber
\mathbb{E}
\left| \cK_{N,\delta}- \cK \right|^2 & = &
\frac{1}{N \delta^2}\Bigg(
c_1 \frac{1}{\kappa^4}  + c_2 \delta \frac{1}{\kappa^3}
+c_3 {\delta}^{2} \frac{1}{\kappa^2} +   c_4\delta^2  + \\
& & \qquad\qquad c_5 \frac{\delta}{\kappa}
+
c_6 \kappa^2\delta^2 + c(\delta\kappa)\Bigg)
\nonumber \\ 
\nonumber && +
\frac{1}{N^2 \delta^2}\Bigg(
d_1 \frac{1}{\kappa^4}  + d_2 \delta \frac{1}{\kappa^3}
+ d_3 {\delta}^{2} \frac{1}{\kappa^2} +  
\\
& & \qquad \qquad 
d_4\delta^2  + d_5 \frac{\delta}{\kappa}
+
d_6 \kappa^2\delta^2 + d(\delta\kappa)\Bigg), \label{e:variance_shear}
\end{eqnarray}
where the constants $\{c_i, \, d_i ; i=1, \dots 6\}$ can be calculated
explicitly and $c(\delta \kappa), \, d( \delta \kappa)$ converge to a
constant exponentially quickly in the limit as $\delta \kappa
\rightarrow + \infty$. From the above formula we immediately deduce
that
\begin{eqnarray*}
\lim_{\kappa \rightarrow 0}\mathbb{E} \left| \cK_{N, \delta}-\cK \right|^2 =   0 
\end{eqnarray*}
provided that~\eqref{e:scaling_shear} holds, together with $N \sim
\kappa^{-2 - \eps}$, $\eps >0$. Furthermore, under these assumptions
on $\delta$ and $N$ we have that
\begin{equation}\label{e:conv_rate_2}
\lim_{\kappa \rightarrow 0} \kappa^{-\eps} \mathbb{E} \left| \cK_{N, \delta}-\cK \right|^2 = \mbox{const}.
\end{equation}
This example shows that Theorem~\ref{thm:l2_unrescaled} is not
sharp. Some details of the calculation of the first two moments of the quadratic
variation for the time independent two-dimensional shear flow are presented
in Appendix~\ref{sec:shear}.


\section{Numerical experiments}
\label{sec:numerics}
In this section we illustrate the results of the previous sections
with some numerical experiments, and we investigate some modifications
to the eddy diffusivity estimator which we shall describe below. The
purpose of the numerical experiments that we have performed is to
investigate the following issues: 
\begin{enumerate}
\item The performance of the estimator~\eqref{e:quad_variation} for
  the eddy diffusivity as a function of the sampling rate for flows
  with different streamline topologies.
\item Whether an appropriate averaging procedure can reduce the
  variance of the estimator.
\item The performance of the estimator~\eqref{e:quad_variation} for
  the eddy diffusivity as a function of the sampling rate for the 
  rescaled problem.
\item The performance of the estimator~\eqref{e:quad_variation} in
  the presence of measurement noise.
\end{enumerate}
The main conclusions from our numerical experiments can be summarised
as follows:
\begin{enumerate}
\item The variance of the estimator as well as the optimal sampling
rate depend crucially on the streamline topology of the velocity
field.
\item Shift averaging (see below) marginally reduces the variance due
  to multiscale error of the estimator, whereas box averaging (also
  see below) introduces extra bias into the estimator.
\item There is an optimal sampling rate for the estimator applied to
  the rescaled problem, but even when using the optimal sampling rate
  the variance of the estimator can be very large.
\item When the data is subject to measurement noise then subsampling
  is necessary, even in the absence of multiscale error. Appropriate
  averaging can reduce the variance due to measurement error.
\end{enumerate}

%
%
\subsection{The Estimators}

We are given a time series of Lagrangian observations of
length $T$, sampled at a constant rate $\Delta t$. The number of
observations is $N = T/\Delta t$. Our goal is to estimate the eddy diffusivity 
using the quadratic variation~\eqref{e:quad_variation}
\begin{equation}\label{e:quad}
\cK_{N,\delta} = \frac{1}{2 N \delta} \sum_{n=0}^{N-1} \big( x_{n+1}- x_n
\big) \otimes \big( x_{n+1}- x_n \big),
\end{equation}
We will consider both the unrescaled~\eqref{e:sde} as well as the rescaled problems
~\eqref{e:rescaled}. The results presented in Sections~\eqref{sec:resc} 
and~\eqref{sec:small_kappa} suggest that subsampling at an appropriate rate is necessary 
in order to estimate the eddy diffusivity correctly, using Lagrangian observations. In the
numerical experiments presented in this section we will take the sampling rate to scale 
either with $\kappa$ (for the unrescaled problem) or with $\epsilon$ (for the rescaled 
problem), according to the results presented in Theorems~\ref{thm:l2_unrescaled} 
and~\eqref{thm:rescaled}:
$$
\delta \sim \kappa^\alpha, \quad \mbox{or} \quad \delta \sim \eps^{\alpha},
$$
for some appropriate exponent $\alpha$.

Even if we use~\eqref{e:quad} with $\delta$ chosen optimally, the
resulting estimator is clearly not optimal since we are using only a
very small portion of the available data.  Furthermore, the variance
of~\eqref{e:quad_variation} with subsampled data can be enormous, in
particular when $\kappa \ll 1$ or $\eps \ll 1$. One may attempt to
reduce the bias and variance in the estimator by making use of all the
data.  In particular, it is reasonable to expect that subsampling
combined with averaging over the data might lead to a more efficient
estimator of the eddy diffusivity with reduced bias in comparison to
the estimator~\eqref{e:quad}. This methodology was applied
in~\cite{AitMykZha05a, AitMykZha05b} in order to estimate the
integrated stochastic volatility in the presence of market
microstructure noise (observation error).

The most natural way of averaging over the data is by splitting the
data into $N_B$ bins of size $\delta$ with $\delta N_B = N$ and to
perform a local averaging over each bin. We use the notation
$$
x^j_n :=x ((n-1) \delta + (j-1) \Delta t), \quad n=1, \dots N_B, \;
 j=1, \dots J, \quad JN_B = N,
$$
 for the $j$-th observation in the $n$-th bin. $J=\delta/\Delta t$ is
 the number of observations in each bin. The maximum likelihood estimator
\eqnref{e:quad_variation} is then computed using the averaged values
\[
\bar{x}_n = \frac{1}{J}\sum_{j=1}^J x^j_n,
\]
leading to the {\bf box-averaged} estimator:
\begin{equation}
\label{box averaging}
  \cK_{N_B,\delta}^b = \frac{1}{2 N_B \delta} \sum_{n=0}^{N_B-1} \left(
\frac{1}{J}\sum_{j=1}^J x^j_{n+1}
-
\frac{1}{J}\sum_{j=1}^J x^j_n
  \right) \otimes \left(
\frac{1}{J}\sum_{j=1}^J x^j_{n+1}
-
\frac{1}{J}\sum_{j=1}^J x^j_n
\right).
\end{equation}
A second averaging technique, proposed in \cite{AitMykZha05b, AitMykZha05a} 
to remove the effects of market microstructure noise,
is to compute a series of estimators, each using a different
observation from each bin, and then to compute the average. This is
the {\bf shift-averaged} estimator:
\begin{equation}
\label{shift averaging}
  \cK_{N_B,\delta}^s =
\frac{1}{J}\sum_{j=1}^J
\frac{1}{2 N_B \delta}
\sum_{n=0}^{N_B-1} \left(
x^j_{n+1}
-
 x^j_n
  \right) \otimes \left(
 x^j_{n+1}
-
 x^j_n
\right).
\end{equation}
In all of the tests the box-averaged and shift-averaged estimators
were obtained using values from every single timestep. Throughout this
section, we only consider the component of the eddy diffusivity along
the direction of the shear, since only that component is modified by 
the flow.
%
%
\subsection{The Velocity Fields}

The numerical experiments were performed using the following four
different idealized divergence-free velocity fields in two dimensions:
\begin{enumerate}
\item { The two-dimensional shear flow}:
\begin{equation}\label{e:shear}
{\bf v}({\bf x}) = (0,\sin(x)),
\end{equation}
for which the eddy diffusivity is
is~\cite{kramer}
\[
\cK = \kappa + \frac{1}{2\kappa}.
\]
\item { The periodically-modulated two-dimensional shear flow}:
\begin{equation}\label{e:shear_t}
{\bf v}({\bf x},t) = (0,\sin(x)\sin(\omega t)),
\end{equation}
with $\omega>0$, for which the eddy diffusivity~\cite{wiggins}
is
\[
\cK = \kappa + \frac{1}{4(\omega + \kappa^2)}.
\]
\item { The stochastically-modulated two-dimensional shear flow}:
\begin{equation}\label{e:ou_shear}
{\bf v}({\bf x},t) = (0,\eta(t)\sin(x)),
\end{equation}
where $\eta(t)$ is an Ornstein-Uhlenbeck process obtained from
the equation
\[
\dot{\eta}(t) = -\alpha \eta(t) + \sqrt{2\sigma}\dot{\beta},
\]
and where $\beta$ is a one-dimensional Brownian motion.  The
 eddy diffusivity is
\begin{equation}\label{e:Deff_ou_shear}
\cK = \kappa + \frac{\sigma}{2(\kappa+\alpha)\alpha}.
\end{equation}
The calculation of the eddy diffusivity for this velocity field is presented in Appendix \ref{sec:deff_ou_shear}.
\item { The Taylor-Green flow}:
\begin{equation}\label{e:TG}
v(x,t) = \nabla^{\bot} \psi_{TG}(x,y), \quad \psi_{TG}(x,y) = \sin(x)\sin(y).
\end{equation}
There is no closed formula for the eddy diffusivity for this flow, but
it is well known~\cite{childress,childress1, childress2,Fann01, Kor04} that the
eddy diffusivity  is isotropic and that
$$
\cK = c^*\kappa^{1/2}, \quad \kappa \ll 1
$$
with a formula for the prefactor $c^*$. For this case we obtain a numerical
approximation to the eddy diffusivity $\cK$ using the spectral
method described in~\cite{MajMcL93, thesis}.
\end{enumerate}
We remark that, whereas in the case of the time independent shear flow the eddy diffusivity
becomes singular as $\kappa \rightarrow 0$, in all other examples the eddy diffusivity
vanishes in the zero molecular diffusion limit. The rate of convergence of $\cK$ to $0$ is
different for the velocity fields~\eqref{e:shear_t},~\eqref{e:ou_shear} and the
Taylor-Green flow~\eqref{e:TG}. From Theorem~\ref{thm:l2_unrescaled} we expect that
the different scaling of the eddy diffusivity with $\kappa$ should manifest itself
in the scaling of the optimal subsampling rate with $\kappa$.\footnote{The analysis
presented in Section~\ref{sec:small_kappa} applies only to time-independent velocity
fields, but can be easily generalized to cover the case of time dependent velocity fields.
In fact, for the velocity fields~\eqref{e:shear_t} and~\eqref{e:ou_shear} we
can analyze directly the quadratic variation without appeal to a general theory. See
Appendix~\eqref{sec:shear}. }

\subsection{Results}

Numerical solutions to \eqnref{e:lagrange} were obtained for each of
these cases using the Euler-Maruyama method with a very small timestep
to remove the effects of numerical discretisation error. The estimator
\eqnref{e:quad_variation} was then computed for each numerical
trajectory and compared with the correct value. In the case of the
averaged estimators we used all the data in each bin to compute the
averages. These calculations were repeated for 1000 realisations of
the trajectory with different Brownian motions, and mean and standard
deviations for the estimator values were computed.
%
%
\subsubsection{The Unrescaled Process}
\label{s:unrescaled_nums}

Figure \ref{shearbars} shows the results of the three estimators
applied to the shear flow for various values of $\delta$ with an
interval width $T=1000$, from which the number of bins $N_B=T/\delta$
for the averaged estimators can be computed. As is consistent with
equation \eqnref{e:kappa_lim}, the maximum likelihood estimator
\eqnref{e:quad_variation} underestimates the eddy diffusivity, and
converges to the small-scale diffusivity $\kappa$ for small $\delta$.
For larger $\delta$, the mean value of the maximum likelihood
estimator approaches the correct value of the eddy diffusivity, but
the standard deviation of the estimator becomes large, indicating a
large variance which means that the probability of accurately
estimating the correct value is small. In comparison, the
shift-averaged estimator does not improve the bias by much and the
variance is only reduced slightly. The box-averaged estimator
increases the bias in the estimator in the sense that it substantially
underestimates the eddy diffusivity.

Figure \ref{modbars} shows the same information for the
periodically-modulated shear flow with modulation frequency $\omega$.
The small $\delta$ limit is again consistent with equation
\eqnref{e:kappa_lim}, and the mean of the estimator increases to a
maximum which is well above the correct value, before decreasing
again, with increasing standard-deviations for large values of
$\delta$. The shift-averaging again shows very little improvement in
either the bias or the variance; the box-averaging reduces the mean
towards zero in all cases.

Figure \ref{OUbars} shows the same information for the OU-modulated
shear flow with parameters $\alpha=1$, $\sigma=0.1$. The results for
the maximum likelihood estimator indicate an optimum value for
$\delta$ which corresponds with a maximum of the mean, however the
standard deviation increases monotonically with $\delta$. There is a
small improvement in the bias and standard deviation for the
shift-averaging, and the box-averaging produces a mean which is less
than the small-scale diffusivity $\kappa$ for all values of $\delta$.

Figure \ref{TGbars} shows the same information for the Taylor-Green
flow. We observe, as is consistent with our theory, that there does
seem to be an optimum sampling rate, but the variance is large near
the optimal rate, similar to the other cases.

%
\subsubsection{The Rescaled Problem}

We then repeated all of these computations for the rescaled problem
\eqnref{e:rescaled} with $\epsilon=0.1$. Figures \ref{eps01shearbars},
\ref{eps01modbars}, \ref{eps01OUbars}, and \ref{eps01TGbars} show the
results for the shear flow, the periodically-modulated shear flow, the
OU-modulated shear flow and the Taylor-Green flow respectively. Each
of these flows showed that there is an optimal sampling rate for which
the mean of the maximum likelihood estimator is close to the correct
value, and that the standard deviation is not too large at this
sampling rate, although the standard deviation increases for large
sampling rates. This illustrates the result of theorem
\ref{thm:rescaled}: the mean of the maximum likelihood estimator
converges to the correct value as $\epsilon\to 0$ and the variance
converges to zero as the subsampling rate $\delta$ converges to zero.

%
%
\subsubsection{The Effect of Observation Noise}
%
%
In this section we consider the combined effect of the multiscale
structure and of measurement noise; measurement noise is included
using equation~\eqref{e:noise}. The experiments of section
\ref{s:unrescaled_nums} were repeated, with $\theta=0.1$.  Figures
\ref{eps01shearbars}, \ref{eps01modbars}, \ref{eps01OUbars}, and
\ref{eps01TGbars} show the results for the shear flow, the
periodically-modulated shear flow, the OU-modulated shear flow and the
Taylor-Green flow respectively.  These results confirm
equation~\eqref{e:unresc_noise} in showing that the expectation of the
estimators tends to infinity as $\delta$ tends to 0 for non-zero
$\theta$. This means that it becomes necessary to subsample even if
there is no multiscale error. The results also show that for
$\theta=0.1$, the multiscale error dominates the variance of the
estimator when subsampling is applied. The shift-averaging technique
is effective at removing the variance due to measurement error, but
not the variance due to multiscale error.

\section{Conclusions}
\label{sec:conclusions}

The problem of estimating the eddy diffusivity from noisy Lagrangian
observations was studied in this paper. Apart from the direct
relevance of our findings to the problem of the accurate
parameterisation of the effects of small scales in oceanic models, we
believe that this work is also a step towards the development of
efficient methods for data-driven coarse graining. Problems similar to
the ones considered in this paper have been studied in the context of
data assimilation. For example, one might fit data from the full dynamics
(i.e. the primitive equations) to the quasi-geostrophic
equation which is a reduced model which is obtained from the full
dynamics after averaging, in the limit as the Rossby number $Ro$ goes
to $0$. Our results suggest that great care has to be
taken when fitting data to a reduced model which is not compatible
with the data at all scales. This is particularly the case when the
reduced model is obtained through a singular limit such as $Ro
\rightarrow 0$.

In this paper, we considered this problem for a class of velocity
fields (divergence-free, smooth, periodic in space and either steady
or modulated in time) for which it can be shown rigorously that a
parameterisation of the Lagrangian trajectories exists, in terms of an
eddy diffusivity tensor. For this class of flows, it was shown, by
means of analysis and numerical experiments, that subsampling is
necessary in order to be able to estimate the eddy diffusivity from
Lagrangian observations. It was also shown that the optimal sampling
rate depends on the topological properties of the velocity field.

Parameter estimation methods that combine subsampling with averaging
of the data (defined as shift averaging and box averaging) were also
proposed. It was shown that shift averaging is very efficient in
reducing the effects of observation error, but only slightly reduces
the variance of the estimator. It appears that the shift-averaging
technique is only useful for removing measurement error (or
microstructure noise in the case of econometrics) and not for reducing
the multiscale error, as defined in the introduction. On the other
hand, box averaging leads to a biased estimator, even when the optimal
sampling rate is used. This should not be surprising, since in the
trivial case where the velocity field vanishes (\emph{i.e.} pure
Brownian motion with diffusivity $\kappa$), the expectation of the box
averaged estimator is $\kappa/J$ where $J$ is the number of points per
bin. On the other hand, for the same problem, the expectation of the
shift averaged estimator is $\kappa$. 

For efficient accurate coarse graining it is necessary to develop
estimators which can deal with the multiscale error more
efficiently. Appropriate averaging over the data appears to be an
important ingredient of such an estimator. An alternative method has
been proposed in \cite{CVE06a} based on the reconstruction of the
generator of the observed Markov process; methods that combine
subsampling and averaging with this approach are currently being
developed.

We believe that our conclusions extend to more general types of
velocity fields. For example, one can carry out the analysis and
numerical experiments presented in this paper using the class of
incompressible Gaussian random velocity fields that were considered in
\cite{CC99}. This appears to be a general class of models to
consider since one can obtain velocity fields with any chosen energy
spectrum. The regularity of such velocity fields should definitely
play an important role in the statistical inference procedure.

Clearly the calculation of the optimal sampling rate from the data is
crucial for our approach. It appears that frequency domain techniques
are more suitable for addressing this issue, and this will be
investigated in subsequent publications.

{\bf Acknowledgements.} The authors are particularly grateful to
A.M. Stuart and P.R. Kramer for their very careful reading of an
earlier draft of the paper and for many useful suggestions and
comments.

\bibliography{../bibtex_files/mybib}

\def\cprime{$'$} \def\cprime{$'$} \def\cprime{$'$} \def\cprime{$'$}
  \def\cprime{$'$} \def\cprime{$'$} \def\cprime{$'$}
  \def\Rom#1{\uppercase\expandafter{\romannumeral #1}}\def\u#1{{\accent"15
  #1}}\def\Rom#1{\uppercase\expandafter{\romannumeral #1}}\def\u#1{{\accent"15
  #1}}\def\cprime{$'$} \def\cprime{$'$} \def\cprime{$'$} \def\cprime{$'$}
  \def\cprime{$'$} \def\cprime{$'$} \def\cprime{$'$}
\begin{thebibliography}{ASMZ05b}

\bibitem[ASMZ05a]{AitMykZha05b}
Y.~Ait-Sahalia, P.~A. Mykland, and L~Zhang.
\newblock How often to sample a continuous-time process in the presence of
  market microstructure noise.
\newblock {\em Rev. Financ. Studies}, 18:351--416, 2005.

\bibitem[ASMZ05b]{AitMykZha05a}
Y.~Ait-Sahalia, P.~A. Mykland, and L~Zhang.
\newblock A tale of two time scales: Determining integrated volatility with
  noisy high-frequency data.
\newblock {\em J. Amer. Stat. Assoc.}, 100:1394--1411, 2005.

\bibitem[AM90]{ma:mmert}
{\sc M.~Avellaneda and A.~J. Majda}, {\em Mathematical models with exact
  renormalization for turbulent transport}, Comm. Pure Appl. Math., 131 (1990),
  pp.~381--429.

\bibitem[AM91]{AvelMajda91}
{\sc M.~Avellaneda and A.~J. Majda}, {\em An integral representation and bounds
  on the effective diffusivity in passive advection by laminar and turbulent
  flows}, Comm. Math. Phys., 138 (1991), pp.~339--391.

\bibitem[BSGMO98]{sb:emfde}
{\sc S.~Bauer, M.~S. Swenson, A.~Griffa, A.~J. Mariano, and K.~Owens}, {\em
  Eddy-mean flow decomposition and eddy-diffusivity estimates in the tropical
  pacific ocean. 1. {M}ethodology}, J. Geophys. Res., 103 (1998),
  pp.~30855--30871.

\bibitem[BSG02]{sb:emfde2}
{\sc S.~Bauer, M.~S. Swenson, and A.~Griffa}, {\em Eddy mean flow decomposition
  and eddy diffusivity estimates in the tropical {P}acific {O}cean: 2.
  {R}esults}, J. Geophys. Res., 107 (2002), p.~3154.
\newblock doi:10.1029/2000JC000613.

\bibitem[Bat99]{bhatta_2}
R.~Battacharya.
\newblock Multiscale diffusion processes with periodic coefficients and an
  application to solute transport in porous media.
\newblock {\em The Annals of Applied Probability}, 9(4):951--1020, 1999.

\bibitem[BGW89]{bhatta_1}
R.~N. Bhattacharya, V.K. Gupta, and H.F. Walker.
\newblock Asymptotics of solute dispersion in periodic porous media.
\newblock {\em SIAM J. APPL. MATH}, 49(1):86--98, 1989.

\bibitem[BLP78]{lions}
A.~Bensoussan, J.-L. Lions, and G.~Papanicolaou.
\newblock {\em Asymptotic analysis for periodic structures}, volume~5 of {\em
  Studies in Mathematics and its Applications}.
\newblock North-Holland Publishing Co., Amsterdam, 1978.

\bibitem[BM02]{BeMc02}
P.~S. Berloff and J.C. McWilliams.
\newblock Material transport in oceanic gyres. {Part II: H}ierarchy of
  stochastic models.
\newblock {\em J. Phys. Oceanogr.}, 32:797--830, 2002.

\bibitem[BM03]{BeMc03}
P.~S. Berloff and J.C. McWilliams.
\newblock Material transport in oceanic gyres. {Part III: R}andomized
  stochastic models.
\newblock {\em J. Phys. Oceanogr.}, 33:1416--1445, 2003.

\bibitem[BR80]{BasRao80}
I.V. Basawa and B.L.S.~Prakasa Rao.
\newblock {\em Statistical inference for stochastic processes}.
\newblock Academic Press Inc. [Harcourt Brace Jovanovich Publishers], London,
  1980.

\bibitem[CDRS09]{CDRS09}
S.~Cotter, M.~Dashti, J.C. Robinson, and A.M. Stuart.
\newblock Data assimilation problems in fluid mechanics: {B}ayesian formulation
  in function space.
\newblock {\em Reprint}, 2009.

\bibitem[Chi79]{childress2}
S.~Childress.
\newblock Alpha effect in flux ropes and sheets.
\newblock {\em Phys. Earth and Planet. Int.}, 20:172--180, 1979.

\bibitem[CKRZ97]{ConstKiselRyzhZl06}
P.~Constantin, A.~Kiselev, L.~Ryzhik, and A.~Zlatos.
\newblock Diffusion and mixing in fluid flow.
\newblock {\em Preprint}, 1997.

\bibitem[CS89]{childress}
S.~Childress and A.M. Soward.
\newblock Scalar transport and alpha-effect for a family of cat's-eye flows.
\newblock {\em J. Fluid Mech.}, 205:99--133, 1 989.

\bibitem[CC99]{CC99}
R.A. Carmona and F. Cerou.
\newblock Transport by incompressible random velocity fields:
simulations \& mathematical conjectures.
\newblock in {\em Stochastic partial differential equations: six perspectives},
Math. Surveys Monogr. 64, 153--181, Amer. Math. Soc. 1999. 

\bibitem[CX97]{carmona}
R.A. Carmona and L.~Xu.
\newblock Homogenization theory for time-dependent two-dimensional
  incompressible gaussian flows.
\newblock {\em The Annals of Applied Probability}, 7(1):265--279, 1997.

\bibitem[CVE06a]{CVE06a}
D.T. Crommelin and E.~Vanden-Eijnden.
\newblock Reconstruction of diffusions using spectral data from timeseries.
\newblock {\em Commun. Math. Sci.}, 4(3):651--668, 2006.

\bibitem[CVE06b]{CVE06b}
D. T. Crommelin and E. Vanden-Eijnden.
\newblock Fitting timeseries by continuous-time Markov chains: a
quadratic programming approach.
\newblock {\em J. Comput. Phys.}, 217(2):782--805, 2006.

\bibitem[Fan02]{Fann01}
A.~Fannjiang.
\newblock Time scales in homogenization of periodic flows with vanishing
  molecular diffusion.
\newblock {\em J. Differential Equations}, 179(2):433--455, 2002.

\bibitem[Fig94]{haf:eredd2}
{\sc H.~A. Figueroa}, {\em Eddy resolution versus eddy diffusion in a double
  gyre {GCM}. part ii: Mixing of passive tracers}, J. Phys. Oceanogr., 24
  (1994), pp.~387--402.

\bibitem[FO94]{haf:eredd}
{\sc H.~A. Figueroa and D.~B. Olson}, {\em Eddy resolution versus eddy
  diffusion in a double gyre {GCM}. part i: The lagrangian and eulerian
  description}, J. Phys. Oceanogr., 24 (1994), pp.~371--386.

\bibitem[GKS04]{GKS04}
D. Givon, R. Kupferman and A. Stuart.
\newblock Extracting macroscopic dynamics: model problems and algorithms.
\newblock {\em Nonlinearity}, 17(6):R55=R127, 2004.

\bibitem[GOPR95]{Gr+95}
A.~Griffa, K.~Owens, L.~Piterbarg, and B.~Rozovskii.
\newblock Estimates of turbulence parameters from {L}agrangian data using a
  stochastic particle model.
\newblock {\em Journal of Marine Research}, 53(3):371--401, 1995.

\bibitem[Hei03]{Heinz03}
S.~Heinze.
\newblock Diffusion-advection in cellular flows with large {P}eclet numbers.
\newblock {\em Arch. Ration. Mech. Anal.}, 168(4):329--342, 2003.

\bibitem[HP08]{HP07}
M.~Hairer and G.~A. Pavliotis.
\newblock From ballistic to diffusive behavior in periodic potentials.
\newblock {\em J. Stat. Phys.}, 131(1):175--202, 2008.

\bibitem[HKDS07]{IKDS07}
I. Horenko, R. Klein, S. Dolaptchiev and C. Sch\"utte.
\newblock Automated generation of reduced stochastic weather models.
I. Simultaneous dimension and model reduction for time series
analysis.
\newblock {\em Multiscale Model. Simul.}, 6(4):1125--1145, 2007.

\bibitem[HS08]{HS08}
I. Horenko and C. Sch{\"u}tte.
\newblock Likelihood-based estimation of multidimensional {L}angevin
models and its application to biomolecular dynamics.
\newblock {\em Multiscale Model. Simul.}, 7(2)::731--773, 2008.

\bibitem[Kor04]{Kor04}
L.~Koralov.
\newblock Random perturbations of 2-dimensional {H}amiltonian flows.
\newblock {\em Probab. Theory Related Fields}, 129(1):37--62, 2004.

\bibitem[KS91]{KSh91}
I.~Karatzas and S.E. Shreve.
\newblock {\em Brownian {M}otion and {S}tochastic {C}alculus}, volume 113 of
  {\em Graduate Texts in Mathematics}.
\newblock Springer-Verlag, New York, second edition, 1991.

\bibitem[Kut04]{Kut04}
Y.A. Kutoyants.
\newblock {\em Statistical inference for ergodic diffusion processes}.
\newblock Springer Series in Statistics. Springer-Verlag London Ltd., London,
  2004.

\bibitem[MBW96]{wiggins}
I.~Mezic, J.F. Brady, and S.~Wiggins.
\newblock Maximal effective diffusivity for time-periodic incompressible fluid
  flows.
\newblock {\em SIAM J. APPL. MATH}, 56(1):40--56, 1996.

\bibitem[McL98]{McL98}
R.M. McLaughlin.
\newblock Numerical averaging and fast homogenization.
\newblock {\em J. Statist. Phys.}, 90(3-4):597--626, 1998.

\bibitem[MK99]{kramer}
A.J. Majda and P.R. Kramer.
\newblock Simplified models for turbulent diffusion: {T}heory, numerical
  modelling and physical phenomena.
\newblock {\em Physics Reports}, 314:237--574, 1999.

\bibitem[MM93]{MajMcL93}
A.J. Majda and R.M. McLaughlin.
\newblock The effect of mean flows on enhanced diffusivity in transport by
  incompressible periodic velocity fields.
\newblock {\em Stud. Appl. Math.}, 89(3):245--279, 1993.

\bibitem[Pav02]{thesis}
G.~A. Pavliotis.
\newblock {\em Homogenization Theory for Advection -- Diffusion Equations with
  Mean Flow, Ph.D Thesis}.
\newblock Rensselaer Polytechnic Institute, Troy, NY, 2002.

\bibitem[Pit02]{Pi02}
L.~I. Piterbarg.
\newblock The top {L}yapunov exponent for a stochastic flow modeling the upper
  ocean turbulence.
\newblock {\em SIAM Journal on Applied Mathematics}, 62(3):777--800, 2002.

\bibitem[PPS08a]{PapPavSt08}
T.~Papavasiliou, G.A. Pavliotis, and A.M. Stuart.
\newblock Maximum likelihood estimation for multiscale diffusions.
\newblock {\em Preprint}, 2008.

\bibitem[PPS08b]{PavlPokStu08}
G.A. Pavliotis, Y.~Pokern, and A.M. Stuart.
\newblock Parameter estimation for multiscale diffusions: an overview.
\newblock {\em Preprint}, 2008.

\bibitem[PS07]{PavlSt06}
G.~A. Pavliotis and A.~M. Stuart.
\newblock Parameter estimation for multiscale diffusions.
\newblock {\em J. Stat. Phys.}, 127(4):741--781, 2007.

\bibitem[PS08]{PavlSt08}
G.A. Pavliotis and A.M. Stuart.
\newblock {\em Multiscale methods}, volume~53 of {\em Texts in Applied
  Mathematics}.
\newblock Springer, New York, 2008.
\newblock Averaging and homogenization.

\bibitem[PSZ07]{PavlStuZyg07}
G.~A. Pavliotis, A.~M. Stuart, and K.~C. Zygalakis.
\newblock Homogenization for inertial particles in a random flow.
\newblock {\em Commun. Math. Sci.}, 5(3):507--531, 2007.

\bibitem[SC90]{childress1}
A.M. Soward and S.~Childress.
\newblock Large magnetic {R}eynolds number dynamo action in spatially periodic
  flow with mean motion.
\newblock {\em Proc. Roy. Soc. Lond. A}, 331:649--733, 1990.

\bibitem[ARGO06]{ARGO06}
The~ARGO team.
\newblock Five years of progress, five decades of potential.
\newblock promotional brochure, March 2006.

\bibitem[VGRM04]{mv:otsms}
{\sc M.~Veneziani, A.~Griffa, A.~M. Reynolds, and A.~J. Mariano}, {\em Oceanic
  turbulence and stochastic models from subsurface {L}agrangian data for the
  northwest {A}tlantic {O}cean}, J. Phys. Oceanogr., 34 (2004), pp.~1884--1906.

\bibitem[Wig05]{Wi05}
S.~Wiggins.
\newblock The dynamical systems approach to {L}agrangian transport in oceanic
  flows.
\newblock {\em Annual Review of Fluid Mechanics}, 37:295--328, 2005.

\end{thebibliography}
\bibliographystyle{plain}
%
%
\appendix

\section{Derivation of Formula~\eqref{e:Deff_ou_shear}}
\label{sec:deff_ou_shear}
In this appendix we derive the formula for the effective diffusivity for the OU-modulated
shear flow~\eqref{e:ou_shear}. Homogenization problems for Gaussian incompressible
velocity fields that are given in terms of an Ornstein-Uhlenbeck process have been
considered in~\cite{carmona, PavlStuZyg07}. The results presented in these papers imply
that
$$
\lim_{\eps \rightarrow 0} \eps y(t/\eps^2) = \sqrt{2 \cK} W(t),
$$
weakly on $C([0,T];\R)$ where $W(t)$ is a standard one-dimensional Brownian motion and
\begin{equation}\label{e:Deff_ou_shear1}
\cK = \kappa + \kappa \|\partial_x \phi \|^2_{L^2(X; \rho)} +
\sigma \|\partial_\eta \phi \|^2_{L^2(X; \rho)}.
\end{equation}
We have used the notation $X:= (2 \pi \T)^2 \times \R$, and $\phi$ and $\rho$
are the unique solutions of the equations
\begin{subequations}\label{e:phi_rho}
\begin{equation}\label{e:phi}
-\cL \phi = \eta \sin(x), \quad \int_{X} \phi \rho \, dX = 0,
\end{equation}
\begin{equation}\label{e:rho}
-\cL^* \rho = 0,  \quad \int_{X} \rho \, dX = 1.
\end{equation}
\end{subequations}
We have used the notation $d X = dx dy d \eta$ and $\cL$ is the generator of the Markov
process restricted on $X$:
$$
\cL = \eta \sin(x) \partial_y + \kappa \partial_x^2 + \kappa \partial_y^2 -\alpha \eta
\partial_\eta + \sigma  \partial_\eta^2.
$$
$\cL^*$ denotes the $L^2(X)$-adjoint, i.e. the Fokker-Planck operator. We can easily solve
equations~\eqref{e:phi} and~\eqref{e:rho} to obtain
$$
\rho \, dx dy d \eta = \frac{1}{Z} e^{-\frac{\alpha \eta^2}{2 \sigma^2}} \, dx dy d \eta,
\quad Z = 4 \pi^2 \sqrt{\frac{2 \pi \sigma}{\alpha}}
$$
and
$$
\phi(x,y,\eta) = \frac{1}{\kappa +\alpha} \eta \sin(x).
$$
Consequently:
\begin{eqnarray*}
\|\partial_x \phi \|_{(L^2;\rho)}^2 & = & \frac{1}{(\kappa +\alpha)^2} Z^{-1} \int_X \eta^2
(\cos(x))^2 \rho \, dX  \\ & = & \frac{\sigma}{2 \alpha} \frac{1}{(\kappa +\alpha)^2}
\end{eqnarray*}
and
\begin{eqnarray*}
\|\partial_\eta \phi \|_{(L^2;\rho)}^2 & = & \frac{1}{(\kappa +\alpha)^2} Z^{-1}
\int_X (\sin(x))^2 \rho \, dX  \\ & = & \frac{1}{2(\kappa +\alpha)^2}.
\end{eqnarray*}
Upon inserting the above two formulas in~\eqref{e:Deff_ou_shear1} we
obtain~\eqref{e:Deff_ou_shear}.

\section{The two-dimensional shear flow}
\label{sec:shear}

In this appendix we study in more detail the problem of estimating the
eddy diffusivity from Lagrangian observations for a class of
two-dimensional shear flows. Throughout this appendix we only consider
the eddy diffusivity along the direction of the shear.  The flows that
we will consider are of the form
\begin{equation}\label{e:shear_appendix}
v(x,y,t) = (0, \eta(t) f(x)),
\end{equation}
where $f(x)$ is a smooth periodic function and $\eta(t)$ is either a constant,
a smooth periodic function of time or a stochastic process, e.g. the Onrstein-Uhlenbeck process
$$
\frac{d \eta}{d t} = - \alpha \eta + \sqrt{2 \sigma} \frac{d W}{d t}. 
$$
As  it has already been noted in~\cite{ma:mmert,McL98, kramer}, for this class of  velocity fields the Lagrangian equations
can be solved explicitly. In particular, we have that 
\begin{equation}\label{e:y_shear}
y(t) = y(0) + \int_0^t \eta(s) f(x(0) + \sqrt{2 \kappa } W_1(s)) \, ds
+  \sqrt{2\kappa} \,  W_2(t),
\end{equation}
where $W_1(t)$ and $W_2(t)$ are one dimensional independent Brownian motions.
Hence, the formula for the quadratic variation becomes
\begin{equation}
\label{estimator}
\cK_{N,\delta} = \frac{1}{2N\delta}\sum_{n=0}^{N-1}\left(\int_{n \delta}^{(n+1)
 \delta} \eta(s) f(x(0) + \sqrt{2 \kappa } W_1(s)) \, ds
+  \sqrt{2\kappa} \Delta W_2(n\delta)  \right)^2,
\end{equation}
where $\Delta W_2(n\delta) = W_2((n+1)\delta) - W_2(n\delta)$. Since $f(x)$ is a periodic function, the calculation of the statistics of
the quadratic variation can be accomplished by calculating the statistics
of integrals of trigonometric functions of the Brownian motion. This calculation
can be done by using properties of integrals of symmetric functions, that
is functions $f:[n \delta, (n+1)\delta]^d \mapsto \R $ for which $f(t_{\sigma_1}, t_{\sigma_2},
\dots t_{\sigma_d}) = f(t_1, t_2, \dots t_d)$ for all permutations $\sigma$
of $(1,2, \dots d)$. In this way, we can calculate the quadratic variation
as a function of $\kappa$ and $\delta$ in an explicit form. For simplicity we will consider the case $\eta(t) \equiv 1$,  $f(x) = \sin(x)$
and $x(0)= y(0) = 0$. The general case can be treated similarly.

For the velocity field
$$
v(x,y) = (0, \sin(x))
$$ 
We can calculate the expectation of the 22-component of the quadratic
variation equation~\eqref{e:quad_shear}, and hence
prove~\eqref{e:conv_rate}.  Since $W_1(t)$ and $W_2(t)$ are
independent, we immediately deduce that
\begin{eqnarray*}
\mathbb{E}\left[(y_{n+1}-y_n)^2
\right] 
& = & \int_{n\delta}^{(n+1)\delta} \int_{n\delta}^{(n+1)\delta}
\mathbb{E}\left[
\sin(\sqrt{2\kappa}W_1(s_1))
\sin(\sqrt{2\kappa}W_1(s_2))\right]
\dt{s_2}\dt{s_1} + 2\kappa\delta.
\end{eqnarray*}
In order to calculate the integral on the right hand side of the above equation
(which we denote by $S$), we use trigonometric identities together with the
formula for the expectation of the characteristic function of a Gaussian
random variable to obtain
\begin{eqnarray*}
S & = & -\frac{1}{4}
\sum_{\MM{a}\in I}a_1a_2
\int_{n\delta}^{(n+1)\delta} \int_{n\delta}^{(n+1)\delta}
\mathbb{E}\left[e^{i\sqrt{2\kappa}\left(a_1W_1(s_1)+a_2W_1(s_2)\right)}
\right]\dt{s_2}\dt{s_1} 
\\  &=&  -\frac{1}{4}
\sum_{\MM{a}\in I}a_1a_2
\int_{n\delta}^{(n+1)\delta} \int_{n\delta}^{(n+1)\delta}
e^
{-2\kappa\delta\sum_{i,j=1}^2a_ia_j\min(s_i,s_j)}
\dt{s_2}\dt{s_1},
\end{eqnarray*}
where $\MM{a}=(a_1,a_2)$ and $I$ is the index set
$\{-1,1\}\times\{-1,1\} =\{-1,1\}^2$.
The integrand is symmetric in $s_1$ and $s_2$, and, using properties of multiple
integrals of symmetric functions, we can write the above integral in the
form
\begin{eqnarray*}
S &=&  -\frac{1}{8}
\sum_{\MM{a}\in I}a_1a_2
\int_{n\delta}^{(n+1)\delta} \int_{s_1}^{(n+1)\delta}
e^
{-2\kappa\delta\sum_{i=1}^2s_i(a_i^2 + \sum_{i<j}a_ia_j)}
\dt{s_2}\dt{s_1}.
\end{eqnarray*}
Evaluating this formula using Maple gives
\begin{equation}\label{e:shear_exp}
S =  2\kappa\delta + \frac{\delta}{\kappa} +
\frac{1}{2\kappa^2}
\Bigg(-\frac{1}{6}e^{-4\kappa(n+1)\delta}
-\frac{1}{2}e^{-4\kappa n\delta}
+
\frac{2}{3}e^{-\kappa(4n+1)\delta}
+2e^{-\kappa\delta}-2
\Bigg),
\end{equation}
from which~\eqref{e:quad_shear} follows upon summation.

We can also calculate $\mathbb{E} \left| \cK_{N, \delta}-  \cK \right|^2$,
leading to equation~\eqref{e:variance_shear}, and hence~\eqref{e:conv_rate_2}.
We have
\begin{equation}
\label{total variance}
\mathbb{E}\left| \cK_{N,\delta}-\cK \right|^2
=
\mathbb{E}\left| \cK_{N, \delta}\right|^2
-2\mathbb{E}\left(\cK_{N, \delta}\right)\left(\kappa+\frac{1}{2\kappa}\right)
 + \left(\kappa+\frac{1}{2\kappa}\right)^2.
\end{equation}
We have already calculated the expectation of the quadratic variation, and
it remains to compute the second moment. We have
\begin{eqnarray*}
\mathbb{E}
\left|\cK_{\delta}\right|^2
&=& \frac{1}{4N^2\delta^2}
\sum_{n=1}^N\sum_{m=1}^N
\mathbb{E}\left((y^n-y^{n-1})^2(y^m-y^{m-1})^2\right) \\
&=& \frac{1}{4N^2\delta^2}
\sum_{n=1}^N\underbrace{\mathbb{E}\left[(y^n-y^{n-1})^4\right]}_{=S_1^n}
 + \frac{1}{2N^2\delta^2}
\sum_{n=1}^N\sum_{m<n}
\underbrace{\mathbb{E}\left[(y^n-y^{n-1})^2(y^m-y^{m-1})^2\right]}_{=S_2^{nm}}.
\end{eqnarray*}
We shall separately compute these two types of terms, namely the diagonal
terms $S_1^n$ and the off-diagonal terms $S_2^{nm}$.

First we compute $S_1^n$.
\begin{eqnarray*}
S_1^n & = & \underbrace{\mathbb{E}\left[
\left(\int_{n\delta}^{(n+1)\delta}
\sin(\sqrt{2\kappa}W_1(s))\diff{s}\right)^4
\right]}_{S_{11}^n} \\
&  & + 12\kappa\delta\left(\frac{\delta}{\kappa} +
\frac{1}{2\kappa^2}
\Bigg(-\frac{1}{6}e^{-4\kappa(n+1)\delta}
-\frac{1}{2}e^{-4\kappa n\delta}
+
\frac{2}{3}e^{-\kappa(4n+1)\delta}
+2e^{-\kappa\delta}-2
\Bigg)\right) \\ && + 12 \kappa^2 \delta^2,
\end{eqnarray*}
where~\eqref{e:shear_exp} has been used. For the calculation of $S_{11}^n$
we use trigonometric identities, together with the formula for the expectation
of the characteristic function of a Gaussian random variable to obtain
\begin{eqnarray*}
S_{11}^n
& = &
\frac{1}{16} \sum_{\MM{a}\in I}  \prod_{k=1}^4a_k
\int_{s_k=n\delta}^{(n+1)\delta}
e^{-2\kappa
\sum_{i,j=1}^4a_ia_j\min(s_i,s_j)
}\diff{s}_k ,
\end{eqnarray*}
where $\MM{a}=(a_1,a_2,a_3,a_4)$ and $I$ is the indexing set
$\{-1,1\}^4$. The integrand in this multiple integral is a symmetric function,
 and hence we may write
\begin{eqnarray*}
S_{11}^n & = & \frac{3}{2} \sum_{\MM{a}\in I} \prod_{k=1}^4a_k
\int_{s_1=n\delta}^{(n+1)\delta}
\int_{s_2=s_1}^{(n+1)\delta}
\int_{s_3=s_2}^{(n+1)\delta}
\int_{s_4=s_3}^{(n+1)\delta}
e^{-2\kappa
\sum_{i,j=1}^4a_ia_j\min(s_i,s_j)
}\diff{s}_1\diff{s}_2\diff{s}_3\diff{s}_4. 
\end{eqnarray*}
This can be computed using Maple:
\begin{eqnarray*}
S_{11}^n &=& {\frac {1}{26880}}\,{\frac {1}{{\kappa}^{4} \left( {e^{\kappa\,n\delta
}} \right) ^{16} \left( {e^{\kappa\,\delta}} \right) ^{16}}}
\\ &&
\; +{\frac {
261}{64}}\,{\kappa}^{-4}+{\frac {1}{960}}\,{\frac {1}{ \left( {e^{
\kappa\,n\delta}} \right) ^{16}{\kappa}^{4} \left( {e^{\kappa\,\delta}
} \right) ^{4}}}-{\frac {1}{3360}}\,{\frac {1}{ \left( {e^{\kappa\,n
\delta}} \right) ^{16}{\kappa}^{4} \left( {e^{\kappa\,\delta}}
 \right) ^{9}}}-{\frac {45}{16}}\,{\frac {\delta}{{\kappa}^{3}}}
\\ && \;
 -{
\frac {49}{12}}\,{\frac {1}{{\kappa}^{4}{e^{\kappa\,\delta}}}}-{\frac
{1}{2400}}\,{\frac {1}{{\kappa}^{4} \left( {e^{\kappa\,n\delta}}
 \right) ^{4} \left( {e^{\kappa\,\delta}} \right) ^{9}}}+{\frac {1}{
120}}\,{\frac {\delta}{{\kappa}^{3} \left( {e^{\kappa\,n\delta}}
 \right) ^{4} \left( {e^{\kappa\,\delta}} \right) ^{4}}}
\\ && \;
 +{\frac {19}{
24}}\,{\frac {1}{{\kappa}^{4} \left( {e^{\kappa\,n\delta}} \right) ^{4
}}}-{\frac {1}{480}}\,{\frac {1}{ \left( {e^{\kappa\,n\delta}}
 \right) ^{16}{\kappa}^{4}{e^{\kappa\,\delta}}}}-{\frac {229}{288}}\,{
\frac {1}{{\kappa}^{4} \left( {e^{\kappa\,n\delta}} \right) ^{4}{e^{
\kappa\,\delta}}}}
\\ && \;
-5/4\,{\frac {\delta}{{\kappa}^{3}{e^{\kappa\,\delta
}}}}+3/4\,{\frac {{\delta}^{2}}{{\kappa}^{2}}}+{\frac {1}{192}}\,{
\frac {1}{{\kappa}^{4} \left( {e^{\kappa\,\delta}} \right) ^{4}}}
\\ && \;
-{
\frac {5}{12}}\,{\frac {\delta}{{\kappa}^{3} \left( {e^{\kappa\,n
\delta}} \right) ^{4}{e^{\kappa\,\delta}}}}-3/8\,{\frac {\delta}{{
\kappa}^{3} \left( {e^{\kappa\,n\delta}} \right) ^{4}}}+{\frac {7}{
1800}}\,{\frac {1}{{\kappa}^{4} \left( {e^{\kappa\,n\delta}} \right) ^
{4} \left( {e^{\kappa\,\delta}} \right) ^{4}}}+{\frac {1}{768}}\,{
\frac {1}{ \left( {e^{\kappa\,n\delta}} \right) ^{16}{\kappa}^{4}}}
\end{eqnarray*}

After summation, all the terms containing exponentials given rise
to terms which converge to a constant divided by $\delta^2N^2$ faster
than any polynomial power of $\delta\kappa$ as $\kappa\delta\to \infty$.

Next we compute $S_2$. Since $n < m$, the term inside the sum is
\begin{eqnarray*}
\mathbb{E}\left[
(y^{n+1}-y^n)^2(y^{m+1}-y^m)^2
\right]
&=& \mathbb{E}\Bigg[
\left(\int_{n\delta}^{(n+1)\delta}
\sin(\sqrt{2\kappa}W_1(s))\diff{s}
+ \sqrt{2\kappa}\int_{n\delta}^{(n+1)\delta}
\diff{W_2}(s)\right)^2 \\
& & \quad \times
\left(\int_{m\delta}^{(m+1)\delta}
\sin(\sqrt{2\kappa}W_1(s))\diff{s}
+ \sqrt{2\kappa}\int_{m\delta}^{(m+1)\delta}
\diff{W_2}(s)\right)^2
\Bigg]
 \\
& = & \underbrace{\mathbb{E}\Bigg[
\left(\int_{n\delta}^{(n+1)\delta}
\sin(\sqrt{2\kappa}W_1(s))\diff{s}\right)^2
\left(\int_{m\delta}^{(m+1)\delta}
\sin(\sqrt{2\kappa}W_1(s))\diff{s}\right)^2
\Bigg]}_{S_{21}^{nm}} \\
&  & +
2\kappa\delta \Bigg(
2\kappa\delta + \frac{\delta}{\kappa} +
\frac{1}{2\kappa^2}
\Bigg(-\frac{1}{6}e^{-4\kappa(m+1)\delta}
-\frac{1}{2}e^{-4\kappa m\delta}
+ \\
& & \qquad\qquad\qquad\qquad\qquad\qquad\qquad
\frac{2}{3}e^{-\kappa(4m+1)\delta}
+2e^{-\kappa\delta}-2
\Bigg) \Bigg) \\
&  & +
2\kappa\delta \Bigg(
2\kappa\delta + \frac{\delta}{\kappa} +
\frac{1}{2\kappa^2}
\Bigg(-\frac{1}{6}e^{-4\kappa(n+1)\delta}
-\frac{1}{2}e^{-4\kappa n\delta}
+ \\
& & \qquad\qquad\qquad\qquad\qquad\qquad\qquad
\frac{2}{3}e^{-\kappa(4n+1)\delta}
+2e^{-\kappa\delta}-2
\Bigg) \Bigg) \\
&  & \quad +4\kappa^2\delta^2,
\end{eqnarray*}
where~\eqref{e:shear_exp} has been used again. A similar calculation to that
for $S_{11}^n$, making use of the fact that $m>n$ gives
\begin{eqnarray*}
S_{21}^{nm}
& = &
\int_{s_1=n\delta}^{(n+1)\delta}
\int_{s_2=n\delta}^{(n+1)\delta}
\frac{1}{16}
\sum_{\MM{a}\in I}\left(\prod_{k=1}^4a_k\right)
e^{-2\kappa
\left(\sum_{i,j=1}^2a_ia_j\min(s_i,s_j) + 2\sum_{i=1}^2\sum_{j=3}^4a_ia_js_i\right)}
\\
& & \qquad \times
\int_{s_3=m\delta}^{(m+1)\delta}
\int_{s_4=m\delta}^{(m+1)\delta}
e^{-2\kappa
\sum_{i,j=3}^4a_ia_j\min(s_i,s_j)
}\diff{s}_1\diff{s}_2\diff{s}_3\diff{s}_4.
\end{eqnarray*}
The integrand for the two inner integrals, and the integrand for the two
outer integrals are both symmetric functions, and we obtain
\begin{eqnarray*}
S_{21}^{nm}
& = & \frac{1}{4}
\int_{s_1=n\delta}^{(n+1)\delta}
\int_{s_2=s_1}^{(n+1)\delta}
\sum_{\MM{a}\in I}\left(\prod_{k=1}^4a_k\right)
e^{-2\kappa
\left(\sum_{i=1}^2s_i\left(a_i^2+ 2\sum_{i<j<3}a_ia_j\right)
+ 2\sum_{i=1}^2\sum_{j=3}^4a_ia_js_i\right)}
\\
& & \qquad \times\int_{s_3=m\delta}^{(m+1)\delta}
\int_{s_4=s3}^{(m+1)\delta}
e^{-2\kappa
\sum_{i=3}^4s_i\left(a_i^2+ 2\sum_{i<j}a_ia_j\right)
}\diff{s}_1\diff{s}_2\diff{s}_3\diff{s}_4.
\end{eqnarray*}
This can be computing using Maple:
\begin{eqnarray*}
S_{21}^{n m} &=&  -2\,{\frac {1}{{\kappa}^{4}{e^{\delta\,\kappa}}}}+{\frac {1}{2016}}\,{
\frac {1}{{\kappa}^{4} \left( {e^{\kappa\,m\delta}} \right) ^{4}
 \left( {e^{\delta\,\kappa}} \right) ^{16} \left( {e^{\kappa\,n\delta}
} \right) ^{12}}}+{\frac {1}{210}}\,{\frac {1}{{\kappa}^{4} \left( {e^
{\kappa\,m\delta}} \right) ^{4} \left( {e^{\delta\,\kappa}} \right) ^{
6} \left( {e^{\kappa\,n\delta}} \right) ^{12}}}
\\ &  & \;
-{\frac {1}{504}}\,{
\frac {1}{{\kappa}^{4} \left( {e^{\kappa\,m\delta}} \right) ^{4}
 \left( {e^{\delta\,\kappa}} \right) ^{13} \left( {e^{\kappa\,n\delta}
} \right) ^{12}}}
\\ &  & \;
+{\frac {1}{1440}}\,{\frac {1}{{\kappa}^{4} \left( {e
^{\kappa\,m\delta}} \right) ^{4} \left( {e^{\delta\,\kappa}} \right) ^
{4} \left( {e^{\kappa\,n\delta}} \right) ^{12}}}-{\frac {1}{840}}\,{
\frac {1}{{\kappa}^{4} \left( {e^{\kappa\,m\delta}} \right) ^{4}
 \left( {e^{\delta\,\kappa}} \right) ^{9} \left( {e^{\kappa\,n\delta}}
 \right) ^{12}}}-1/12\,{\frac {1}{{\kappa}^{4} \left( {e^{\delta\,
\kappa}} \right) ^{5} \left( {e^{\kappa\,n\delta}} \right) ^{4}}}
\\ &  & \;
+{
\frac {1}{60}}\,{\frac {\delta}{{\kappa}^{3} \left( {e^{\kappa\,m
\delta}} \right) ^{4}}}-1/45\,{\frac {\delta}{{\kappa}^{3} \left( {e^{
\kappa\,m\delta}} \right) ^{4}{e^{\delta\,\kappa}}}}+{\frac {1}{180}}
\,{\frac {\delta}{{\kappa}^{3} \left( {e^{\kappa\,m\delta}} \right) ^{
4} \left( {e^{\delta\,\kappa}} \right) ^{4}}}+1/48\,{\frac { \left( {e
^{\kappa\,n\delta}} \right) ^{4}}{{\kappa}^{4} \left( {e^{\kappa\,m
\delta}} \right) ^{4}}}+
\\ &  & \;
{\frac {17}{2700}}\,{\frac {1}{{\kappa}^{4}
 \left( {e^{\kappa\,m\delta}} \right) ^{4} \left( {e^{\delta\,\kappa}}
 \right) ^{4}}}-{\frac {1}{600}}\,{\frac {1}{{\kappa}^{4} \left( {e^{
\kappa\,m\delta}} \right) ^{4} \left( {e^{\delta\,\kappa}} \right) ^{9
}}}+1/4\,{\frac {1}{{\kappa}^{4} \left( {e^{\kappa\,n\delta}} \right)
^{4}}}-{\frac {7}{12}}\,{\frac {1}{{\kappa}^{4}{e^{\delta\,\kappa}}
 \left( {e^{\kappa\,n\delta}} \right) ^{4}}}
\\ &  & \;
 -1/12\,{\frac { \left( {e^
{\kappa\,n\delta}} \right) ^{4} \left( {e^{\delta\,\kappa}} \right) ^{
3}}{{\kappa}^{4} \left( {e^{\kappa\,m\delta}} \right) ^{4}}}+
\\ &  & \;
1/32\,{
\frac { \left( {e^{\kappa\,n\delta}} \right) ^{4} \left( {e^{\delta\,
\kappa}} \right) ^{4}}{{\kappa}^{4} \left( {e^{\kappa\,m\delta}}
 \right) ^{4}}}-1/36\,{\frac { \left( {e^{\kappa\,n\delta}} \right) ^{
4}}{{\kappa}^{4} \left( {e^{\kappa\,m\delta}} \right) ^{4}{e^{\delta\,
\kappa}}}}+1/18\,{\frac { \left( {e^{\delta\,\kappa}} \right) ^{2}
 \left( {e^{\kappa\,n\delta}} \right) ^{4}}{{\kappa}^{4} \left( {e^{
\kappa\,m\delta}} \right) ^{4}}}
\\ &  & \;
+{\frac {1}{288}}\,{\frac { \left( {e^
{\kappa\,n\delta}} \right) ^{4}}{{\kappa}^{4} \left( {e^{\kappa\,m
\delta}} \right) ^{4} \left( {e^{\delta\,\kappa}} \right) ^{4}}}-1/4\,
{\frac {\delta}{{\kappa}^{3} \left( {e^{\kappa\,n\delta}} \right) ^{4}
}}+1/3\,{\frac {\delta}{{\kappa}^{3}{e^{\delta\,\kappa}} \left( {e^{
\kappa\,n\delta}} \right) ^{4}}}-1/12\,{\frac {\delta}{{\kappa}^{3}
 \left( {e^{\delta\,\kappa}} \right) ^{4} \left( {e^{\kappa\,n\delta}}
 \right) ^{4}}}
 \\ &  & \;
 +1/12\,{\frac {1}{{\kappa}^{4} \left( {e^{\delta\,
\kappa}} \right) ^{4} \left( {e^{\kappa\,n\delta}} \right) ^{4}}}+1/3
\,{\frac {1}{{\kappa}^{4} \left( {e^{\delta\,\kappa}} \right) ^{2}
 \left( {e^{\kappa\,n\delta}} \right) ^{4}}}+{\frac {1}{{\kappa}^{4}
 \left( {e^{\delta\,\kappa}} \right) ^{2}}}+2\,{\frac {\delta}{{\kappa
}^{3}{e^{\delta\,\kappa}}}}
\\ &  & \;
+{\frac {{\delta}^{2}}{{\kappa}^{2}}}-2\,{
\frac {\delta}{{\kappa}^{3}}}+{\kappa}^{-4}-{\frac {1}{360}}\,{\frac {
1}{{\kappa}^{4} \left( {e^{\kappa\,m\delta}} \right) ^{4}{e^{\delta\,
\kappa}} \left( {e^{\kappa\,n\delta}} \right) ^{12}}}
\\ &  & \;
+{\frac {1}{480}}
\,{\frac {1}{{\kappa}^{4} \left( {e^{\kappa\,m\delta}} \right) ^{4}
 \left( {e^{\kappa\,n\delta}} \right) ^{12}}}-{\frac {1}{280}}\,{
\frac {1}{{\kappa}^{4} \left( {e^{\kappa\,m\delta}} \right) ^{4}
 \left( {e^{\delta\,\kappa}} \right) ^{5} \left( {e^{\kappa\,n\delta}}
 \right) ^{12}}}
\\ &  & \;
 +{\frac {1}{672}}\,{\frac {1}{{\kappa}^{4} \left( {e^{
\kappa\,m\delta}} \right) ^{4} \left( {e^{\delta\,\kappa}} \right) ^{
12} \left( {e^{\kappa\,n\delta}} \right) ^{12}}}+{\frac {17}{900}}\,{
\frac {1}{{\kappa}^{4} \left( {e^{\kappa\,m\delta}} \right) ^{4}}}
\\ &  & \;
-{
\frac {1}{72}}\,{\frac { \left( {e^{\delta\,\kappa}} \right) ^{3}}{{
\kappa}^{4} \left( {e^{\kappa\,m\delta}} \right) ^{4}}}-{\frac {161}{
5400}}\,{\frac {1}{{\kappa}^{4} \left( {e^{\kappa\,m\delta}} \right) ^
{4}{e^{\delta\,\kappa}}}}+{\frac {1}{54}}\,{\frac { \left( {e^{\delta
\,\kappa}} \right) ^{2}}{{\kappa}^{4} \left( {e^{\kappa\,m\delta}}
 \right) ^{4}}}
\\ &  & \;
 -{\frac {1}{200}}\,{\frac {1}{{\kappa}^{4} \left( {e^{
\kappa\,m\delta}} \right) ^{4} \left( {e^{\delta\,\kappa}} \right) ^{5
}}}+{\frac {1}{150}}\,{\frac {1}{{\kappa}^{4} \left( {e^{\kappa\,m
\delta}} \right) ^{4} \left( {e^{\delta\,\kappa}} \right) ^{6}}}.
\end{eqnarray*}

After the double summation, all the terms containing exponentials
given rise to terms which converge to a constant multiplied by
$(N-1)/\delta^2N^2$ faster than any polynomial power of $\delta\kappa$ as
$\kappa\delta\to \infty$.

Collecting terms
\begin{eqnarray*}
\mathbb{E}
\left|\cK_{N,\delta}\right|^2
& = &
\frac{1}{4 \kappa^2 \delta^2} - \frac{1}{2 \kappa^3 \delta} + \frac{1}{4
\kappa^2} +1 - \frac{1}{\kappa \delta} + \kappa^2 \\ && +
\frac{1}{N \delta^2}\left(
c_1 \frac{1}{\kappa^4}  + c_2 \delta \frac{1}{\kappa^3}
+c_3 {\delta}^{2} \frac{1}{\kappa^2} +   c_4\delta^2  + c_5 \frac{\delta}{\kappa}
+
c_6 \kappa^2\delta^2 + c(\delta\kappa)\right)
\\ && +
\frac{1}{N^2 \delta^2}\left(
d_1 \frac{1}{\kappa^4}  + d_2 \delta \frac{1}{\kappa^3}
+d_3 {\delta}^{2} \frac{1}{\kappa^2} +   d_4\delta^2  + d_5 \frac{\delta}{\kappa}
+
d_6 \kappa^2\delta^2 + d(\delta\kappa)\right),
\end{eqnarray*}
where the constants $\{c_i, \, d_i ; i=1, \dots 6\}$ can be read from
the above formulas and $c(\delta \kappa), \, d( \delta \kappa)$ converge
exponentially fast to a constant in the limit $\delta \kappa \rightarrow
+ \infty$.

Upon computing the remaining terms in equation~\eqref{total variance} we
notice that all leading order terms are cancelled and we end up with
\begin{eqnarray*}
\mathbb{E}
\left| \cK_{N,\delta}- \cK \right|^2 & = &
\frac{1}{N \delta^2}\left(
c_1 \frac{1}{\kappa^4}  + c_2 \delta \frac{1}{\kappa^3}
+c_3 {\delta}^{2} \frac{1}{\kappa^2} +   c_4\delta^2  + c_5 \frac{\delta}{\kappa}
+
c_6 \kappa^2\delta^2 + c(\delta\kappa)\right)
\\ && +
\frac{1}{N^2 \delta^2}\left(
d_1 \frac{1}{\kappa^4}  + d_2 \delta \frac{1}{\kappa^3}
+d_3 {\delta}^{2} \frac{1}{\kappa^2} +   d_4\delta^2  + d_5 \frac{\delta}{\kappa}
+
d_6 \kappa^2\delta^2 + d(\delta\kappa)\right),
\end{eqnarray*}
which is precisely equation~\eqref{e:variance_shear}.

\end{document}